\pdfoutput=1
\documentclass{article}

\usepackage[a4paper,margin=2.5cm]{geometry}
\usepackage{graphicx}
\usepackage{multicol,multirow}
\usepackage{latexsym}

\usepackage{amsmath,amssymb,amsfonts}
\usepackage{mathrsfs}
\usepackage{amsthm}
\usepackage{array}
\usepackage{bm}
\usepackage{tabularx}
\usepackage{makecell}
\usepackage{booktabs}
\usepackage{appendix}
\usepackage[authoryear]{natbib}
\usepackage[T1]{fontenc}
\usepackage[utf8]{inputenc}
\usepackage{times}
\IfFileExists{sourcesanspro.sty}{\usepackage{sourcesanspro}}{}
\usepackage{newtxmath}
\usepackage{textcomp}
\usepackage{xcolor}
\usepackage{hyperref}
\usepackage{siunitx}
\usepackage{subcaption}

\newcommand{\keywords}[1]{\par\noindent\textbf{Keywords: }#1}

\newcommand{\review}[1]{#1}
\newcommand{\correction}[1]{#1}

\title{Embedding model-form uncertainty in probabilistic calibration of digital twins for bridges}

\author{Daniel Andrés Arcones$^{a,b,*}$,  Martin Weiser$^{c}$, Phaedon-Stelios Koutsourelakis$^{a}$, Jörg F. Unger$^{b}$\\
	\small $^{a}$Technical University of Munich, Garching bei München, Germany \\
	\small $^{b}$Bundesanstalt für Materialforschung und -prüfung, Berlin, Germany \\
	\small $^{c}$Zuse Institute Berlin, Berlin, Germany\\\\
	\small
	\begin{tabular}{@{}l l@{}}
		$^{*}$Corresponding author: &
		Daniel Andrés Arcones; \tt{daniel.andres-arcones@bam.de} \\
	\end{tabular}
}

\begin{document}
\maketitle

\begin{abstract}
	Digital twins of bridges rely on physics-based models to infer full-field structural responses from sparse monitoring data. The reliability of these predictions depends on the calibration of model parameters while accounting for discrepancies between the model and the physical system. Such discrepancies, commonly referred to as model-form uncertainty (MFU) can substantially affect predictive reliability, being therefore essential for trustworthy digital-twin predictions. This work presents a calibration framework that explicitly represents the MFU through a stochastic parameter embedding, providing a general methodology for the uncertainty-aware calibration of physics-based models. The framework is demonstrated using a simplified two-dimensional cross-sectional thermal model of the Nibelungenbrücke calibrated against temperature measurements from its monitoring system. By explicitly accounting for MFU, the proposed methodology enables computationally efficient models to be reliably employed within digital-twin environments while maintaining predictive credibility. The framework introduces a three-step calibration strategy that progressively addresses sources of discrepancy while maintaining a clear distinction between \correction{the part of the model error that is represented by the embedded parameters and can be propagated, the residual model error, and the measurement noise.} Variance decomposition is used to characterize the structure of predictive uncertainty and its propagation to quantities of interest. Remaining discrepancies between predictive distributions and observations are quantified using Kolmogorov-Smirnov metrics \correction{and proper scoring rules} to assess predictive consistency across seasonal conditions. The results demonstrate that explicitly accounting for MFU improves the interpretability and reliability of predictive uncertainties and validate the methodology for calibration of physics-based models supporting digital twins in structural health monitoring.
	\end{abstract}


	\section{Introduction} \label{sec:introduction}
	
	Structural Health Monitoring (SHM) and the digitalization of infrastructure are increasingly enabling the deployment of digital twins for critical assets such as bridges \citep{Farrar2012, Worden2015, Farrar2025}. These digital representations aim to support decision-making by continuously integrating monitoring data with predictive models of structural behaviour. In practice, however, the development of high-fidelity physics-based models for bridges is costly and computationally demanding. As a result, and despite increasing research interest \citep{Mousavi2024}, many digital twin implementations rely heavily on purely data-driven approaches that infer structural behaviour directly from monitoring data \citep{Liu2023, Sakr2024}.
	
	Nevertheless, physics-based models such as finite element (FE) representations remain essential when extrapolation, interpretability, or scenario analysis are required. In engineering practice, however, they are often used only for punctual state assessments, where complex models are developed and refined until the required accuracy is achieved. While effective for isolated analyses, this paradigm is poorly suited to operational digital twins that require continuous updating and interaction with monitoring data streams \citep{Mousavi2024}. Surrogate models are frequently employed to reduce computational cost \citep{SchnellenbachHeld2025}, but their predictive reliability is limited by the training domain and they may suffer from reduced interpretability. \review{Additionally, any modelling assumption, decision and simplification inevitably} introduce discrepancies between predictions and observations that increase as model complexity is reduced. These discrepancies must therefore be explicitly quantified and attributed to their respective sources of uncertainty, in particular to model form uncertainty (MFU).
	
	Thermal effects provide a particularly relevant context in which these issues arise. Temperature variations induce significant stresses and deformations in bridge structures and strongly influence SHM measurements \citep{Kromanis2014, Furtmueller2017, Borah2021, Anastasopoulos2023, Yang2025}. Accurate characterization of the thermal field is therefore essential for interpreting monitored responses and for performing thermal compensation of structural measurements \citep{Kromanis2016, Torzoni2022}. In many practical implementations, thermal compensation is achieved locally by pairing structural sensors with nearby temperature measurements \citep{Eisermann2024}. While effective in some cases, this approach neglects the spatial variability of temperature and limits the ability to propagate thermal effects throughout the structure. A physics-based thermal model enables the reconstruction of the full temperature field, allowing thermal deformations to be estimated at arbitrary locations and for various quantities of interest. Such models, typically based on parametrized FE representations, require careful calibration to ensure that predictions are consistent with observations. Among the available approaches \citep{Ereiz2022}, this study adopts a probabilistic model updating strategy based on maximum likelihood estimation \citep{Kaipio2006}, which naturally enables the quantification of uncertainty after calibration.
	
	This study investigates the calibration of a simplified transient thermal model of the Nibelungenbrücke in Worms, Germany, within the context of the SPP~100+ initiative of the Deutsche Forschungsgemeinschaft (DFG) aimed at prolonging the service life of bridges through monitoring and digitalization \citep{Becks2024,Kang2025a,Kang2025b}. The proposed model is intentionally simplified in order to remain computationally efficient and suitable for operational digital twin applications. As a consequence, structural modelling deficiencies are unavoidable and must be addressed through the explicit quantification of MFU during calibration. Embedded uncertainty representations \citep{Sargsyan2015} provide a systematic way to incorporate model discrepancies directly into the model structure. In contrast to classical discrepancy formulations that attribute model error to the predicted outputs \citep{Kennedy2001}, embedded approaches attribute uncertainty to physically interpretable parameters and allow it to be propagated consistently to predictions and derived quantities of interest (QoI). Although such approaches have recently gained attention in uncertainty quantification research \citep{Huan2017, Pernot2017, AndresArcones2026, Kuppa2026}, their application to real SHM systems remains limited. \review{We propose an integral framework for physics-based digital twins that integrates the quantification of MFU, focusing on the characterization, allocation and propagation of these uncertainties. Its application to thermal models of structures is validated using monitoring data from the real bridge. Such framework can be extended to other classes of systems and enables the propagation to QoIs of the uncertainties introduced during the calibration due to the modelling assumptions.}

	\correction{Building on the embedded formulation of \citet{Sargsyan2015,Sargsyan2019} and its extension to noisy measurements and large discrepancies by \citet{AndresArcones2026}, this work demonstrates its application to the continuous monitoring of a real bridge. A computationally efficient thermal model is calibrated against two years of operational data through a staged procedure that separates sensor geometry, embedded model error and residual noise. The resulting predictive distributions are assessed through a seasonal validation using distributional deviations and proper scoring rules. Particular attention is given to the identifiability and variance attribution of the embedded uncertainty, its distinction from parameter-estimation uncertainty and measurement noise, and its propagation to the cross-sectional thermal curvature as a QoI, which is not directly measured.}

\begin{figure}[!htbp]
		\centering
		\includegraphics[width=\linewidth]{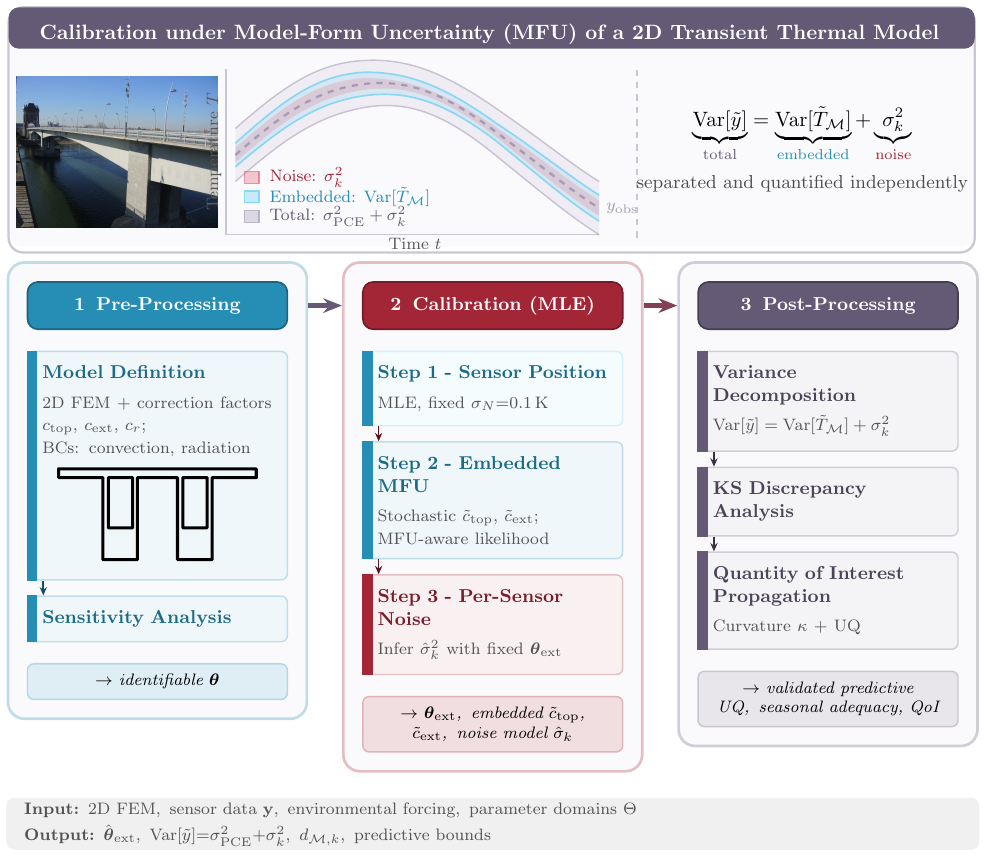}
		\caption{Summary of the framework proposed in this paper. Photo of the Nibelungenbrücke property of Chongjie Kang, TU Dresden}
		\label{fig:summary}
	\end{figure}

	The remainder of this paper is organized as follows. Section~\ref{sec:model} introduces the simplified thermal model and the monitoring data used in this study. Section~\ref{sec:calibration} \correction{ranks the model parameters with a variance-based sensitivity analysis and} describes the three-step calibration framework, including the estimation of sensor positions, the calibration of model parameters with embedded MFU, and the inference of residual noise\correction{, and discusses what the embedding can and cannot separate}. Section~\ref{sec:variance_decomposition} analyzes the predictive variance and its decomposition into different uncertainty sources. Section~\ref{sec:ks_discrepancy} evaluates the agreement between predictions and observations using Kolmogorov-Smirnov discrepancy metrics across different seasonal conditions. \review{Section~\ref{sec:qoi} demonstrates the propagation of the quantified MFU to the cross-sectional curvature as QoI.} A summary of the methodology is provided in Figure~\ref{fig:summary}, and concluding remarks are presented in Section~\ref{sec:conclusion}. \correction{The analysis of the initial condition, the details of the sensitivity analysis and further studies of the embedded representation are collected in the appendices.}
	
	\section{Thermal model and representation of Model-Form Uncertainty} \label{sec:model}
	
	\subsection{Thermal model definition} \label{sub:modelling}

\begin{figure}[!htbp]
		\centering
		\includegraphics[width=\linewidth]{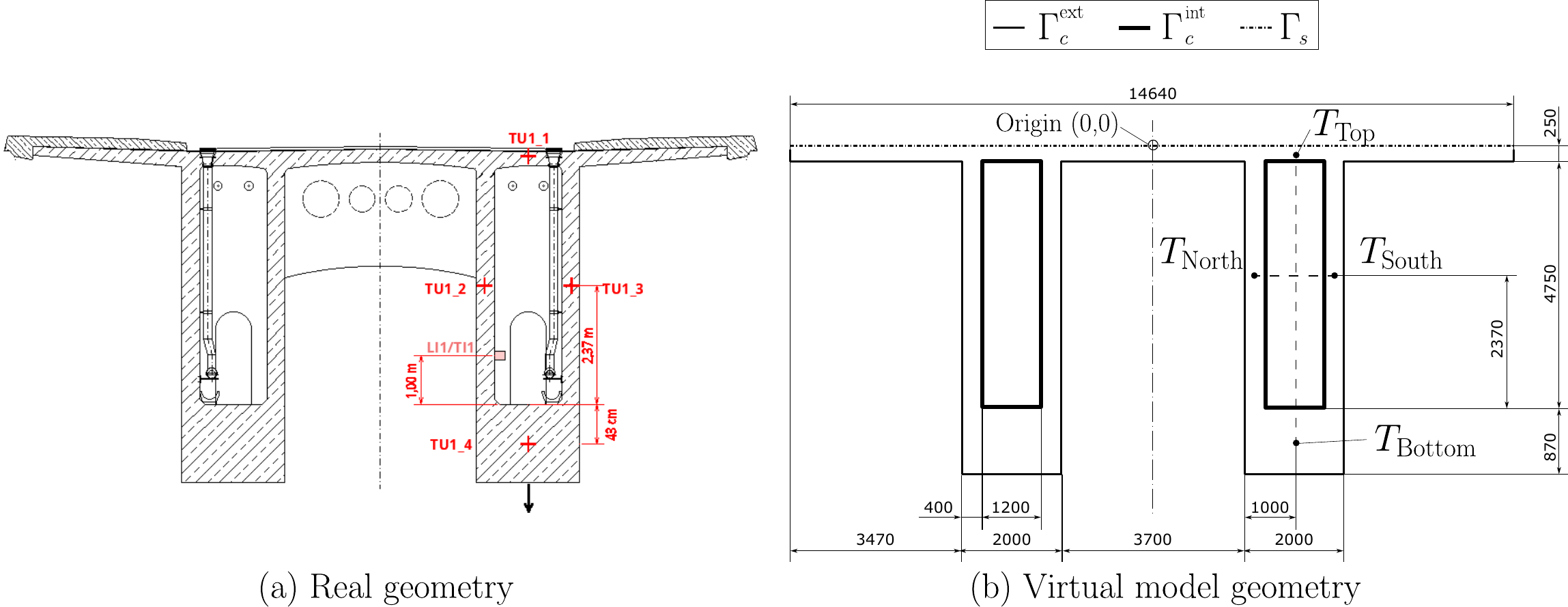}
		\caption{Two-dimensional cross-section of the bridge used in the thermal model. (a) Real geometry as in the monitoring concept drawings (TU Dresden, adapted from \citealp{Kang2024}), (b) virtual model cross-section geometry. In the virtual model, the dimensions are schematic, not at scale. Horizontal positions for $T_{\mathrm{North}}$ and $T_{\mathrm{South}}$ and vertical positions for $T_{\mathrm{Top}}$ and $T_{\mathrm{Bottom}}$ will be inferred due to geometrical discrepancies between real and virtual cross-sections}
		\label{fig:cross-section}
	\end{figure}

	Let $\Omega \subset \mathbb{R}^2$ denote the computational domain representing the bridge cross-section shown in Fig.~\ref{fig:cross-section}. The time interval of interest is $t\in(0,t_f]$. The thermal state of the structure is described by the temperature field $T:\Omega\times(0,t_f]\rightarrow\mathbb{R}$. The thermal model $\mathcal{M}$ is defined as the solution operator that maps the parameter vector $\boldsymbol{\theta}$ and environmental boundary conditions to the temperature field $T(\mathbf{x},t)$ satisfying the transient heat conduction problem
	\begin{align}
		\frac{\partial T}{\partial t} - \nabla\cdot(\alpha\nabla T)
		= 0
		\quad \text{in } \Omega\times(0,t_f],
		\label{eq:heat_eq}
	\end{align}
	together with appropriate boundary and initial conditions, $\alpha$ is the thermal diffusivity. As presented in \citet{Zhu2017}, bridge thermal models are affected by a multitude of complex and interacting boundary conditions that change over time. However, the initial model will be limited to wind-forced convection in the exterior, natural convection in the interior, and short-wave radiation on the top of the deck. The discrepancy generated by the remaining effects will be considered as part of the MFU and quantified through the parameter calibration. Therefore, the boundary $\partial\Omega$ is decomposed as $\partial\Omega = \Gamma_c^{\mathrm{int}} \cup \Gamma_c^{\mathrm{ext}} \cup \Gamma_s$, where $\Gamma_c^{\mathrm{int}}$ denotes internal surfaces exposed to indoor air, $\Gamma_c^{\mathrm{ext}}$ represents external surfaces subject to wind-driven convection, and $\Gamma_s$ denotes external surfaces exposed to solar radiation in addition to convection. The outward unit normal vector to $\partial\Omega$ is denoted by $\mathbf{n}$.	The model is completed with the initial condition
	\begin{align}
		T(\mathbf{x},0) = T_0(\mathbf{x})
		\quad \text{in } \Omega,
	\end{align}
	and the boundary conditions
	\begin{align}
		-\alpha\rho c_p \nabla T \cdot \mathbf{n} &=
		\correction{h_{\mathrm{int}}}
		\left(T - T_\infty^{\mathrm{int}}(t)\right)
		&& \text{on } \Gamma_c^{\mathrm{int}}\times(0,t_f],
		\label{eq:bc_int}
		\\
		-\alpha\rho c_p \nabla T \cdot \mathbf{n} &=
		\correction{c_\mathrm{ext}} h_{\mathrm{ext}}
		\left(T - T_\infty^{\mathrm{ext}}(t)\right)
		&& \text{on } \Gamma_c^{\mathrm{ext}}\times(0,t_f],
		\label{eq:bc_ext}
		\\
		-\alpha\rho c_p \nabla T \cdot \mathbf{n} &=
		\correction{c_\mathrm{top}} h_{\mathrm{ext}}
		\left(T - T_\infty^{\mathrm{ext}}(t)\right)
		-
		c_r \correction{a_s} I_{\mathrm{in}}(t)
		&& \text{on } \Gamma_s\times(0,t_f].
		\label{eq:bc_shortwave}
	\end{align}
	Here $\rho$ denotes the material density and $c_p$ the specific heat capacity. The parameters $h_{\mathrm{int}}$ and $h_{\mathrm{ext}}$ represent the internal and external convective heat transfer coefficients, while $T_\infty^{\mathrm{int}}(t)$ and $T_\infty^{\mathrm{ext}}(t)$ denote the corresponding ambient temperatures. The quantity $I_{\mathrm{in}}(t)$ is the incident shortwave solar irradiance and \correction{$a_s$} the surface absorptivity. The dimensionless coefficients \correction{$c_\mathrm{top}$, $c_\mathrm{ext}$} and $c_r$ are correction factors introduced to account for modelling simplifications in the convection and radiation fluxes which will be inferred from the discrepancy between predictions and observations. \correction{The convective exchange has two correction factors: $c_\mathrm{top}$ acts on the deck top $\Gamma_s$, and $c_\mathrm{ext}$ on all other external surfaces $\Gamma_c^\mathrm{ext}$, i.e. the webs, the cantilevers and the bottom slab. The deck top is separated because it receives the solar input, it is exposed to traffic, and the pavement and asphalt layers on top of the concrete are not modelled, which differentiate it from the other faces. The exchange with the air inside the box girder is modelled by natural convection with $h_\mathrm{int}$ and is not corrected.} A summary table with the model parameters is presented in Table~\ref{tab:model-parameters}.
	
	\begin{table}[!htbp] 
		\centering 
		\caption{Summary of model parameters for the 2D thermal bridge simulation. The parameters to be calibrated are indicated in \textbf{bold} together with their default value.} 
		\label{tab:model-parameters} 
		\begin{tabular}{l l l l} 
			\hline \textbf{Symbol} & \textbf{Description} & \textbf{Value} & \textbf{Units} \\ 
			\hline 
			$\rho$ & Material density & 2400 & kgm$^{-3}$ \\ 
			$c_p$ & Specific heat capacity & 870 & Jkg$^{-1}$K$^{-1}$ \\ 
			$\bm{\alpha}$ & \textbf{Thermal diffusivity} & \textbf{0.8} & \textbf{mm$^2$/s} \\ 
			\correction{$a_s$} & Surface absorptivity & 0.275 & -- \\
			\correction{$\mathbf{c_\mathrm{top}}$} & \correction{\textbf{Convection correction factor, deck top}} & \textbf{1.0} & \textbf{--} \\
			\correction{$\mathbf{c_\mathrm{ext}}$} & \correction{\textbf{Convection correction factor, other external faces}} & \textbf{1.0} & \textbf{--} \\
			$\mathbf{c_r}$ & \textbf{Shortwave correction factor} & \textbf{1.0} & \textbf{--} \\ 
			$h_{\mathrm{int}}$ & Internal heat transfer coefficient & 10 & Wm$^{-2}$K$^{-1}$ \\ 
			$h_{\mathrm{ext}}$ & External heat transfer coefficient & computed from Re, Nu & Wm$^{-2}$K$^{-1}$ \\ 
			$k$ & Thermal conductivity of air & 0.025 & Wm$^{-1}$K$^{-1}$ \\ $L$ & Characteristic length & 4.0& m \\ 
			$\nu$ & Kinematic viscosity of air & \correction{$1.5\times10^{-5}$} & m$^2$/s \\ 
			$\mathrm{Pr}$ & Prandtl number of air & 0.71 & -- \\ 
			$T(\mathbf{x},0)$ & Initial temperature & homogeneous & K \\ 
			$I_{\mathrm{in}}(t)$ & Incident shortwave irradiance & time-dependent & Wm$^{-2}$ \\ 
			$T_\infty^{\mathrm{int}}(t)$ & Internal ambient temperature & time-dependent & K \\ 
			$T_\infty^{\mathrm{ext}}(t)$ & External ambient temperature & time-dependent & K \\ 
			$\Delta t$ & Time step (implicit Euler) & \correction{30} & \correction{min} \\ 
			Mesh & FEM discretization & \correction{2210 vertices, 3654 triangles} & -- \\ \hline 
		\end{tabular} 
	\end{table}
	\newpage
	For the internal heat convective coefficient $h_{\mathrm{int}}=10$ Wm$^{-2}$K$^{-1}$, natural convection is assumed. The external convective coefficient $h_{\mathrm{ext}}$ is determined using standard dimensionless correlations. The Reynolds number is defined as
	\begin{align}
		\mathrm{Re} = \frac{UL}{\nu},
	\end{align}
	where $U$ denotes the wind velocity, $L$ is the characteristic length, and $\nu = \correction{1.5\times10^{-5}}$ m$^2$/s is the kinematic viscosity of air. Because of the complex bridge geometry, the characteristic length is approximated within the range 2.51-7.43 m, corresponding respectively to the hydraulic diameter and the half-chord length of an airfoil-like cross-section, as $L=4.0$ m. The Nusselt number is obtained from the empirical correlation
	\begin{align}
		\mathrm{Nu} = 0.037\,\mathrm{Re}^{4/5}\,\mathrm{Pr}^{1/3},
	\end{align}
	where the Prandtl number for air is $\mathrm{Pr}=0.71$. The external heat transfer coefficient is then evaluated as
	\begin{align}
		h_{\mathrm{ext}} = \frac{\mathrm{Nu}\,k}{L},
	\end{align}
	where $k=0.025$ Wm$^{-1}$K$^{-1}$ is the thermal conductivity of air.
	
	The parameter vector $\boldsymbol{\theta}$ contains the quantities estimated during calibration. Given $\boldsymbol{\theta}$ and the environmental forcing data, the thermal model $\mathcal{M}$ produces the temperature field $T(\mathbf{x},t;\boldsymbol{\theta})$ defined over the computational domain $\Omega$. The model prediction corresponding to sensor $k$ is obtained by evaluating this field at the location of the associated virtual sensor $\mathbf{x}^{\text{virtual}}_k$, i.e., $T_{\mathcal{M}}(t,k;\boldsymbol{\theta})=T(\mathbf{x}^{\text{virtual}}_k,t;\boldsymbol{\theta})$.
	
	The governing equations defining the model $\mathcal{M}$ are solved numerically using the finite element method. The spatial domain $\Omega$ is discretized with an unstructured mesh consisting of triangular elements that approximate the bridge cross-section geometry. The temperature field is represented using continuous second-order polynomial basis functions over the mesh. Time integration of the transient problem is performed using an implicit Euler scheme with a time step $\Delta t = 30$ min, consistent with the temporal resolution of the environmental input data. The numerical discretization produces a sequence of approximate temperature fields from which virtual sensor observations are extracted.

	The bridge structure is geometrically slender and its cross-sectional shape varies slowly along the longitudinal direction. In addition, the available monitoring sensors are sparse and primarily located in a limited number of cross-sections. Under these conditions, the dominant thermal gradients occur across the cross-section rather than along the span, making a two-dimensional approximation appropriate for representing the principal heat transfer mechanisms. The use of a cross-sectional model also provides significant computational advantages when performing repeated simulations during calibration and uncertainty quantification. Increasing the geometric fidelity to a full three-dimensional representation would substantially increase computational cost while providing limited additional information given the available measurements. However, the dimensional simplification introduces additional modelling assumptions, whose effects must be accounted for through the explicit treatment of MFU. The modelling strategy adopted here therefore focuses on quantifying and propagating the error associated with these assumptions. The cross-section geometry used in the model is presented in Figure~\ref{fig:cross-section}.
	
	\subsection{Model-form uncertainty formulation}
	
	In the calibration of physics-based thermal models for bridges, discrepancies between model predictions and observations arise not only from measurement noise and uncertain parameters, but also from structural inadequacies of the model itself. These inadequacies are commonly referred to as model form uncertainty (MFU). A widely adopted statistical framework for representing MFU in the calibration of computational models is the formulation proposed by \citet{Kennedy2001}. In this framework, the observation recorded by sensor $k$ at time $t$, denoted $y(t,k)$, is expressed as
	\begin{equation}
		y(t,k) = T_{\mathcal{M}}(t,k;\boldsymbol{\theta}) + \delta(t,k;\boldsymbol{\theta}) + \varepsilon_k .
	\end{equation}
	
	Here $T_{\mathcal{M}}(t,k;\boldsymbol{\theta})$ denotes the prediction of the thermal model $\mathcal{M}$ evaluated at the observation location and time corresponding to sensor $k$. More formally, $T_{\mathcal{M}}$ can be interpreted as a virtual measurement or observation operator that maps the internal state of the thermal model to the observable quantity recorded by the sensor. The vector $\boldsymbol{\theta}$ contains the model parameters to be calibrated, including the closure constants $c$ introduced to account for unresolved physical processes. The term $\delta(t,k;\boldsymbol{\theta})$ represents the model discrepancy, i.e. the systematic difference between the best achievable model prediction and the true physical response. Finally, $\varepsilon_k$ denotes measurement noise, assumed to follow $\varepsilon_k \sim \mathcal{N}(0,\sigma_k^2)$, where $\sigma_k^2$ is the variance associated with sensor $k$.
	
	The discrepancy term $\delta$ captures effects that remain unresolved even after calibration of the model parameters. In the context of bridge thermal behaviour, such effects may arise from environmental conditions that are not explicitly modelled, variations in the state of the bridge (e.g. aging or damage), unknown or poorly characterised loads, geometric simplifications, or other modelling assumptions. Various statistical representations of $\delta$ have been proposed in the literature, typically relying on stochastic processes such as Gaussian processes or related surrogate models \citep{Kennedy2001, Bayarri2009, Plumlee2017, AndresArcones2023}.
	
	A limitation of the additive discrepancy formulation above is that the discrepancy term is external to the model and therefore cannot be naturally propagated through the governing equations when performing forward predictions under new conditions. This limits the usefulness of the formulation when the calibrated model is intended to operate as part of a predictive digital twin. To address this issue, we adopt an embedded representation of MFU following the framework proposed in \citet{AndresArcones2026} for noisy measurements and potentially large discrepancies, which is based on the formulation from \citet{Sargsyan2019}. In this setting, the discrepancy is incorporated directly into the model parameters as
	\begin{equation}
		y(t,k) = T_{\mathcal{M}}\bigl(t,k;\boldsymbol{\theta}+\delta\bigr) + \varepsilon_k .
	\end{equation}
	
	In contrast to the additive formulation, the embedded discrepancy modifies the model inputs themselves, which allows the resulting uncertainty to propagate consistently through the governing equations. This property is particularly advantageous when predictions are required under conditions different from those used during calibration. Nevertheless, the embedded approach remains inherently limited to the domain explored by the available observations. Since the underlying physical model is not fully closed, the representation does not guarantee that all sources of uncertainty affecting future observations are captured. In practice, additional discrepancies may arise from factors that cannot be explicitly controlled or represented within the modelling framework. To account for these residual effects, the observation noise variance $\sigma_k^2$ is interpreted more broadly as an effective error term that encompasses both measurement noise and unresolved sources of variability that remain outside the model representation.
	
	In the present article, the dominant source of model-form uncertainty is assumed to arise from the representation of heat exchange at the boundaries of the computational domain. The boundary heat fluxes depend on environmental processes that are only partially represented in the model, including the simplified geometric description of the bridge cross-section and the neglect of phenomena such as shadowing effects, variations in solar incidence due to the motion of the sun, cloud coverage, and additional thermal loads induced by traffic. These factors introduce uncertainty in both the incoming and outgoing heat fluxes prescribed through the boundary conditions. To account for these effects, the model introduces the correction factors \correction{$c_\mathrm{top}$, $c_\mathrm{ext}$} and $c_r$ (see Equations~\ref{eq:bc_ext} to \ref{eq:bc_shortwave}), which act as multiplicative modifiers of the boundary heat transfer terms. These parameters serve as closure coefficients that compensate for the aggregated modelling error associated with unresolved environmental and geometric effects.

	\correction{The embedding does not capture the whole model discrepancy. It captures the part of the discrepancy that can be represented by a distribution of the embedded parameters and therefore propagated through the model to other locations, times and quantities. This part is referred to as \textit{embedded model error} in the following. The remaining discrepancy, which no value of the embedded parameters reproduces, is absorbed by the noise term together with the measurement noise, referred to as \textit{residual model error} and \textit{measurement noise} respectively.}

	\subsection{Data and preprocessing}
	Let $\mathcal{Y}$ denote the set of temperature observations used for calibration and validation. The measurements are obtained from the structural monitoring system installed on the \textit{Nibelungenbrücke} bridge \citep{Kang2024,Kang2025,Herrmann2024a}. In this work we consider the temperature sensors located in the cross-section close to the pylon, which provide representative information about the thermal state of the structure. The monitoring system consists of physical (real) temperature sensors embedded in the bridge. These sensors measure the temperature at specific physical locations $\mathbf{x}_{\text{real}}$ within the structure and belong to the permanent bridge instrumentation system. The acquisition system records the measurements with a sampling rate of 10~Hz. According to the manufacturer specifications, the measurement uncertainty of the sensors is below $0.1^\circ$C. The spatial coordinates, installation characteristics, and functional description of all physical sensors considered in this study are summarized in Table~\ref{tab:sensor_description}. 
		
	\begin{table}[!htbp]
		\centering
		\caption{Summary of sensors from the monitoring system that will be used in the thermal model calibration. Wind speed data was obtained from OpenMeteo API hourly historical data at the given position.}
		\label{tab:sensor_description}
		\begin{tabular}{l l c c l}
			\hline
			\textbf{ID} & \textbf{Measurement variable} & \textbf{Units} & \shortstack{\textbf{Measurement}\\\textbf{frequency}} & \textbf{Position} \\
			\hline
			TA & Air temperature (outside) & °C & 10 Hz & Above the deck \\
			TI & Air temperature (inside) & °C & 10 Hz & In the box, near the pier \\
			KS & Radiation (short wave) & Wm$^{-2}$ & 10 Hz & Above the deck \\
			WS & Wind speed & m/s & Hourly & 49° 37' N, 8° 22' E\\
			TU1\_1 & Structure's temperature Top & °C & 10 Hz & In the box, near the pier \\
			TU1\_2 & Structure's temperature North & °C & 10 Hz & In the box, near the pier \\
			TU1\_3 & Structure's temperature South & °C & 10 Hz & In the box, near the pier \\
			TU1\_4 & Structure's temperature Bottom & °C & 10 Hz & In the box, near the pier \\
			\hline
		\end{tabular}
	\end{table}
	
	The dataset $\mathcal{Y}$ is constructed from the monitoring data after a preprocessing stage. Raw data are first cleaned within the bridge monitoring infrastructure following the procedures described in \citet{Eisermann2024}, which include automated acquisition, validation and data fusion. For the present study, the cleaned data are subsequently subsampled from the original acquisition frequency of 10~Hz to a lower temporal resolution of 30 min appropriate for the slow thermal dynamics of the bridge. The resulting dataset contains synchronized time series of structural and environmental measurements that are used to construct the calibration and validation sets. These environmental measurements are extended with external weather forecast records or predictions \citep{AndresArcones2023a}. In particular, the wind speed will be generated from OpenMeteo API \citep{Zippenfenig2024} historical archive data. As the available data is hourly and the timestep of interest is 30 min, the wind speed values are interpolated to generate the missing entries in the dataset. Wind direction could be also recalled from the same source, but it will be assumed that the wind is perpendicular to the bridge direction for simplicity. 
	
	For comparison with \review{the real system}, a corresponding set of \emph{virtual sensors} is defined \review{at the numerical thermal model}. Virtual sensors represent model-based observations extracted from the numerical solution at selected spatial locations $\mathbf{x}_{\text{virtual}}$. Their purpose is to reproduce, as closely as possible, the measurements recorded by the real monitoring system while remaining consistent with the spatial discretization and geometric representation of the computational model. In general, the sensor coordinates $\mathbf{x}_{\text{real}} \neq \mathbf{x}_{\text{virtual}}$, as the virtual sensor locations are defined within the simplified cross-sectional geometry and finite element mesh, which do not exactly reproduce the physical geometry and sensor mounting positions of the real structure. As a consequence, even in the absence of parameter uncertainty, a discrepancy between real and virtual sensor observations is expected. This is particularly clear for the sensor close to the top surface, as the asphalt of the road over the bridge is not modelled, which creates a difference in the thickness of the structure at that point. A summary of the real and virtual sensor counterparts is presented in Table~\ref{tab:real_vs_virtual_sensors}. The real sensor positions are projected to the closest point in model as the starting point and reference for the calibration.
	
	\begin{table}[!htbp]
		\centering
		\caption{Correspondence between physical monitoring sensors and virtual sensors defined in the numerical model. The coordinates represent the planned locations of the physical sensors, with the origin of coordinates at the center of the deck surface, as indicated in Figure~\ref{fig:cross-section}.}
		\label{tab:real_vs_virtual_sensors}
		\begin{tabular}{c c c}
			\hline
			\shortstack{\textbf{Real}\\$y(t,k)$} & \shortstack{\textbf{Virtual}\\$T_{\mathcal{M}}(t,k;\boldsymbol{\theta})$} & \shortstack{\textbf{Real sensor position} \\ $(x,y)$ [m]}\\
			\hline
			TU1\_1 & $T_\mathrm{Top}$ & (2.85, -0.15) \\
			TU1\_2 & $T_\mathrm{North}$ & \correction{(2.05, -2.63)} \\
			TU1\_3 & $T_\mathrm{South}$ & \correction{(3.65, -2.63)} \\
			TU1\_4 & $T_\mathrm{Bottom}$ & (2.85, -5.43) \\
			\hline
		\end{tabular}
	\end{table}
	
	The main dataset that will be used in this work for calibrating the model are the observations captured from May 15$^\text{th}$, 2024 at \correction{01:00:00} to June 30$^\text{th}$, 2024 at \correction{01:00:00 (UTC)}. After downsampling, this dataset contains 2209 entries with a timestep $\Delta t$ of 30 min and 9 \correction{measurements} per entry, one per sensor indicated in \correction{Table~\ref{tab:sensor_description}}. \correction{The first 800 timesteps, approximately the first two weeks of data, are the burn-in of the transient model described below and are disregarded for calibration and predictions;} therefore this will be referred as the \textit{June 2024} dataset. An additional multi-year dataset will be used for the seasonal validation of the model, spanning from June 15$^\text{th}$, 2023 at \correction{01:00:00} to \correction{August 31$^\text{st}$, 2025 at 23:30:00 (UTC)} and encompassing \correction{38830} analogous entries. \correction{The shortwave irradiance sensor (KS) of the monitoring system has been faulty since September 2025. Because the irradiance is a boundary input of the model, the predictions after that date are not representative, and the seasonal validation is therefore restricted to the period before the fault.} The processed datasets used in this study are publicly available through a Zenodo repository (see Data availability statement) to ensure reproducibility of the results.
	
	\correction{The initial temperature field $T(\mathbf{x},0)$ of the bridge is unknown. The simulations therefore start from a homogeneous field, which is first evolved for 100 steps with the thermal loads of the first entry of the dataset; the measured loads are then applied from entry 100 onwards and the following 700 steps are discarded, so that the predictions start at entry 800. For homogeneous initial fields between $-3.15$~°C and $26.85$~°C, the largest difference between the temperatures predicted at the sensors is at most $2\times10^{-4}$~°C after the 700 discarded steps, more than two orders of magnitude below the measurement uncertainty of 0.1~°C, so the predictions do not depend on the choice of the initial field.}
	
	\section{Parameter calibration}
	\label{sec:calibration}
	\subsection{Parameter screening}
	\label{sec:sensitivity}
	\correction{Before the calibration, the model parameters are ranked by their influence on the predicted temperatures. The ranking motivates the staged calibration described below and the choice of the parameter that carries the embedded MFU. It is obtained from a variance-based sensitivity analysis: the first- and total-order Sobol' indices \citep{Sobol2001} of the predictions $Y=\lbrace T_\mathrm{Bottom},T_\mathrm{Top}, T_\mathrm{North}, T_\mathrm{South}\rbrace$ of the full thermal model for June~2024 are estimated with the Saltelli sampling scheme \citep{Saltelli2002}, as implemented in \texttt{SALib} \citep{Usher2016}. The parameters are sampled from uniform distributions over the calibration domains of Table~\ref{tab:inference_comparison_no_hr}, so the indices reflect the relative importance of the parameters within these domains.}
	
\begin{figure}[!htbp]
		\centering
		\includegraphics[width=\textwidth]{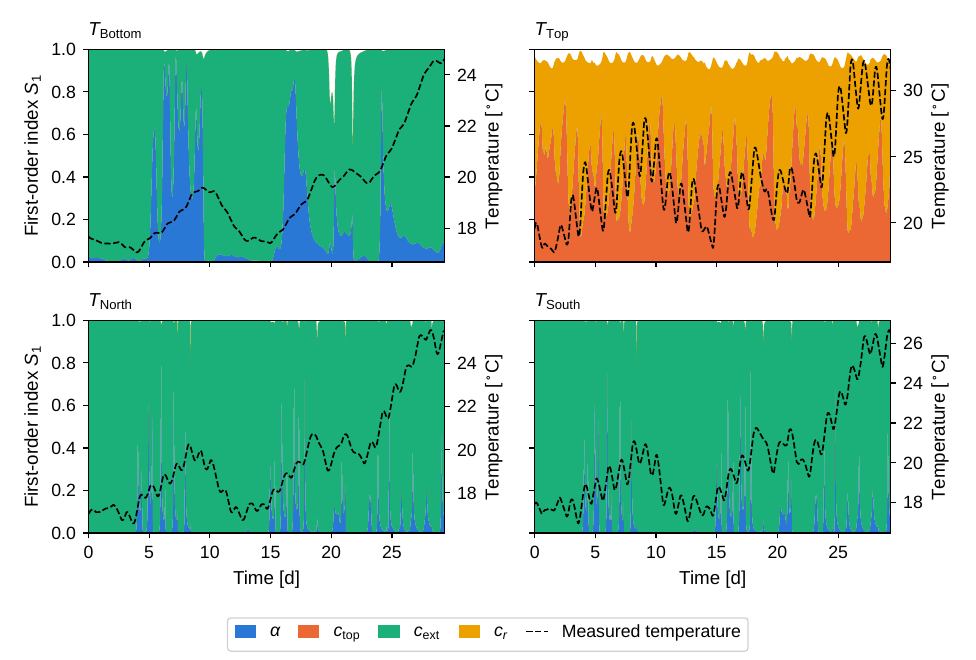}
		\caption{Comparative timeline of the first Sobol indices for June 2024 in the configuration without sensor positions}\label{fig:s1_timeline}
	\end{figure}

	The first-order Sobol' indices $S_1$ are shown in Figure~\ref{fig:s1_timeline} as time-resolved stack plots for the main model parameters and for each temperature sensor. Sensor position parameters strongly dominate the sensitivity of the predictions and have therefore been excluded from the figure to improve readability. The total-order indices $S_T$ are nearly identical to the first-order indices\correction{ at $T_\mathrm{Bottom}$, $T_\mathrm{North}$ and $T_\mathrm{South}$}, indicating that interaction effects between parameters are limited and that the model response is largely driven by individual parameter contributions\correction{; at $T_\mathrm{Top}$, the two deck-top factors interact (see below)}.   
	
	Among the remaining parameters, the convection \correction{factor of the external faces $c_\mathrm{ext}$} dominates the predictions at the $T_\mathrm{Bottom}$, $T_\mathrm{North}$, and $T_\mathrm{South}$ sensors \correction{(time-averaged first-order indices 0.82, 0.95 and 0.95), whereas the predictions at $T_\mathrm{Top}$ depend only on the two factors acting on the deck top, $c_\mathrm{top}$ and $c_r$ (0.42 and 0.53)}. Since convective heat exchange represents the primary heat input to the system, this behaviour is expected and highlights the importance of accurately representing the convective closure term. Consequently, \correction{the two convection factors are} natural candidates to carry an embedded stochastic extension in order to represent model-form uncertainty associated with the convection model. \correction{The shortwave factor $c_r$ has the largest index at $T_\mathrm{Top}$, but it acts on the same boundary as $c_\mathrm{top}$, and the total-order indices of the two (0.55 and 0.58) exceed their first-order indices, implying an interaction between them. A second embedded factor on the deck top, observed by a single sensor, would not be separable from the first. The deck-top exchange is therefore only embedded through $c_\mathrm{top}$, and $c_r$ is estimated without a stochastic extension.} The value ranges for \correction{$c_\mathrm{top}$ and $c_\mathrm{ext}$ have} been chosen arbitrarily large due to the unknown effects. In the ideal case, they would remain around 1.0, but a lower limit of 0.01 would mostly eliminate the effect of their respective heat inputs on the structure, while the upper limit would reflect an exposure 4 times larger than expected. The chosen ranges present the belief at the time of the calibration and condition the resulting sensitivity analysis.
	
	The thermal diffusivity $\alpha$ shows a limited influence and primarily affects $T_\mathrm{Bottom}$ during specific time intervals. Its sensitivity may also be partially inflated due to the relatively broad sampling range used in the analysis, motivated by existing literature \citep{Nagy2018}. When sensor position parameters are included in the analysis, the relative importance of $\alpha$ decreases substantially compared with the convection \correction{factors}, suggesting potential identifiability limitations. For this reason, $\alpha$ is not considered a suitable candidate for stochastic embedding, as its influence is both sensor-specific and restricted to a limited portion of the observations.

	\subsection{Methodology} \label{sub:bayesian_calibration}
	This section presents the calibration framework applied to model $\mathcal{M}$, whose response $T_\mathcal{M}(t,k;\boldsymbol{\theta})$ is compared against temperature measurements $y(t,k)$ collected at $N_k$ sensors over $N_t$ time steps. The goal of calibration is to infer the unknown parameter vector $\boldsymbol{\theta} \in \Theta \subseteq \mathbb{R}^d$ of $d$ parameters from the available data $\mathcal{Y} = \{y(t,k)\}$, while appropriately quantifying the associated uncertainties. The calibration is structured in three sequential steps, each targeting a distinct source of uncertainty: the geometric uncertainty in the sensor positions, the physical parameters governing the model response, and the sensor-specific noise levels.
	
	\paragraph{Maximum Likelihood Estimator for parameter calibration}
	We adopt a statistical model that relates the model predictions to the observed data through a sensor-specific additive noise assumption
	\begin{equation}
		y(t,k) = T_\mathcal{M}(t,k;\boldsymbol{\theta}) + \varepsilon_k, \quad \varepsilon_k \sim \mathcal{N}(0,\sigma_k^2),
		\label{eq:noise_model}
	\end{equation}
	where $\sigma_k^2$ is the variance of the measurement error at sensor $k$, assumed independent across sensors and constant in time. Under this assumption, the conditional distribution of each observation is
	\begin{equation}
		p\left(y(t,k) \mid \boldsymbol{\theta}, \sigma_k^2\right) = \frac{1}{\sqrt{2\pi\sigma_k^2}} \exp\left(-\frac{\bigl[y(t,k) - T_\mathcal{M}(t,k;\boldsymbol{\theta})\bigr]^2}{2\sigma_k^2}\right).
	\end{equation}
	Assuming conditional independence across time steps and sensors, the likelihood function $\mathcal{L}(\boldsymbol{\theta},\left\lbrace\sigma_k^2\right\rbrace\mid\mathcal{Y})$ is defined as the joint probability of the observed data given the parameters
	\begin{align}
		\mathcal{L}\left(\boldsymbol{\theta}, \left\lbrace\sigma_k^2\right\rbrace \mid \mathcal{Y}\right)
		&= \prod_{k=1}^{N_k}\prod_{t=1}^{N_t} p\left(y(t,k)\mid\boldsymbol{\theta},\sigma_k^2\right)\\
		&= \prod_{k=1}^{N_k}\left(2\pi\sigma_k^2\right)^{-N_t/2}
		\exp\left(-\frac{1}{2\sigma_k^2}\sum_{t=1}^{N_t}\bigl[y(t,k)-T_\mathcal{M}(t,k;\boldsymbol{\theta})\bigr]^2\right).
		\label{eq:likelihood}
	\end{align}
	\review{The assumption of independence in time \correction{is a working assumption. The residuals of thermal models are strongly correlated in time: for the calibrated models of this study, the lag-one autocorrelation of the residuals at $\Delta t=30$~min exceeds 0.99 at all sensors, the autocorrelation after one day is 0.45--0.80, and the integrated autocorrelation time $\tau$ is 18 to 150 steps, which corresponds to an effective number of independent residuals $n_\mathrm{eff}=N_t/\tau$ of 9 to 77 out of $N_t=1409$ per sensor in the training month. A likelihood that models this correlation would require a model of the residual process itself. The independent likelihood is kept, and its consequences are accounted for when required. Treating the $N_t$ residuals as independent overstates the information in the data by a factor of about $\tau$: the curvature of the log-likelihood is too large by about $\tau$, so the standard errors derived from it (the square roots of the diagonal of the inverse Hessian, i.e. the parametric uncertainty of the fitted values) are too small by about $\sqrt{\tau}$. We therefore multiply them by $\sqrt{\tau}$ when comparing them with the embedded spread; this is equivalent to using $n_\mathrm{eff}$ instead of $N_t$. The significance of the Kolmogorov-Smirnov tests of Section~\ref{sec:ks_discrepancy} is overstated in the same way.}} For numerical stability and computational convenience, we work with the log-likelihood function
	\begin{equation}
		\ell\left(\boldsymbol{\theta}, \left\lbrace\sigma_k^2\right\rbrace \mid \mathcal{Y}\right)
		= -\frac{N_t}{2}\sum_{k=1}^{N_k}\ln(2\pi\sigma_k^2)
		- \sum_{k=1}^{N_k}\frac{1}{2\sigma_k^2}\sum_{t=1}^{N_t}\bigl[y(t,k) - T_\mathcal{M}(t,k;\boldsymbol{\theta})\bigr]^2.
		\label{eq:log_likelihood}
	\end{equation}
	The Maximum Likelihood Estimator (MLE) $\left(\hat{\boldsymbol{\theta}}_{\text{MLE}}, \{\hat{\sigma}_{k,\text{MLE}}^2\}\right)$ is obtained by solving the following optimization problem:
	\begin{equation}
		\left(\hat{\boldsymbol{\theta}}_{\text{MLE}}, \{\hat{\sigma}_{k,\text{MLE}}^2\}\right)
		= \underset{(\boldsymbol{\theta},\left\lbrace\sigma_k^2\right\rbrace)\in\Theta\times(\mathbb{R}^+)^{N_k}}{\arg\max}
		\;\ell\left(\boldsymbol{\theta},\left\lbrace\sigma_k^2\right\rbrace\mid\mathcal{Y}\right).
		\label{eq:mle_problem}
	\end{equation}
	Since $T_\mathcal{M}(t,k;\boldsymbol{\theta})$ is in general a nonlinear, potentially non-convex function of $\boldsymbol{\theta}$, the optimization problem in Eq.~\eqref{eq:mle_problem} is solved numerically. We employ a gradient-free optimizer, specifically the Nelder-Mead algorithm \citep{Nelder1965} available in the \texttt{scipy} Python package \correction{\citep{Virtanen2020}}. Valid domain ranges are prescribed for the parameters $\boldsymbol{\theta}$ that prescribe $\ell\left(\boldsymbol{\theta}, \left\lbrace\sigma_k^2\right\rbrace \mid \mathcal{Y}\right)=-\infty$ if violated.
	
	The standard formulation assumes that $\mathcal{M}$ is a deterministic map. To account for MFU, we adopt the embedded approach of \citet{Sargsyan2019}, in which a subset of the parameters is promoted to random variables as implemented in \citet{AndresArcones2026}. \review{The extended parameter vector is partitioned as $\boldsymbol{\theta}_\mathrm{ext}=\lbrace\boldsymbol{\theta},\boldsymbol{\theta}_\delta\rbrace$, where $\boldsymbol{\theta}_\delta$ parameterizes the distribution of the stochastic parameter vector $\tilde{\boldsymbol{\theta}}\sim\pi(\cdot|\boldsymbol{\theta},\boldsymbol{\theta}_\delta)=\pi(\cdot|\boldsymbol{\theta}_\mathrm{ext})$ on the parameter space $\Theta$. Structural model discrepancy is thus implicitly absorbed into the variability of the stochastic model response $\tilde{T}_\mathcal{M}(t,k;\tilde{\boldsymbol{\theta}})$, making the observation model consistent with Eq.~\eqref{eq:noise_model}.}
	
	To propagate $\tilde{\boldsymbol{\theta}}$ through the model, a Polynomial Chaos Expansion (PCE) is constructed \citep{Sudret2021}. For fixed $\boldsymbol{\theta}_\mathrm{ext}$ governing the distribution of $\tilde{\boldsymbol{\theta}}$, the stochastic response is approximated as
	\begin{equation}
		\tilde{T}_\mathcal{M}(t,k;\tilde{\boldsymbol{\theta}}) \approx \sum_{\correction{\mathbf{j}} \in \mathcal{A}} \correction{u_{\mathbf{j}}(t,k)\,\Psi_{\mathbf{j}}}\left(\tilde{\boldsymbol{\theta}}\right),
		\label{eq:pce}
	\end{equation}
	where $\{\Psi_{\correction{\mathbf{j}}}\}$ is a multivariate polynomial basis orthonormal with respect to \correction{the density of the germ from which $\tilde{\boldsymbol{\theta}}$ is obtained}, selected following Askey's scheme \citep{Xiu2002}, $\correction{\mathbf{j}} \in \mathbb{N}_0^d$ is a multi-index, $\mathcal{A}$ is the finite truncation set, and $\correction{u_{\mathbf{j}}(t,k)} \in \mathbb{R}$ are the PCE coefficients computed via Gaussian quadrature. The orthonormality condition $\langle\Psi_{\correction{\mathbf{j}}},\Psi_{\correction{\mathbf{j}}}\rangle = 1$ yields closed-form expressions for the first two moments of the stochastic response:
	\begin{align}
		m_\mathrm{PCE}(t,k;\boldsymbol{\theta}_\mathrm{ext}) &= \mathbb{E}\left[\tilde{T}_\mathcal{M}(t,k;\tilde{\boldsymbol{\theta}})\right] \approx \correction{u_{\mathbf{0}}(t,k)}, \label{eq:pce_mean}\\
		\sigma^2_\mathrm{PCE}(t,k;\boldsymbol{\theta}_\mathrm{ext}) &= \mathrm{Var}\left[\tilde{T}_\mathcal{M}(t,k;\tilde{\boldsymbol{\theta}})\right] \approx \sum_{\correction{\mathbf{j}} \neq \mathbf{0}} \correction{u_{\mathbf{j}}^2(t,k)}. \label{eq:pce_var}
	\end{align}
	\correction{In this work, the two embedded parameters $\tilde c_\mathrm{top}$ and $\tilde c_\mathrm{ext}$ introduced in Section~\ref{sub:calibration_results} are independent and lognormal, and each is written as a function of a standard normal germ, $\tilde c_j=\exp(\mu_{\ln,j}+\sigma_{\ln,j}\xi_j)$ with $\xi_j\sim\mathcal{N}(0,1)$, $j\in\lbrace\mathrm{top},\mathrm{ext}\rbrace$, so that the basis consists of products of Hermite polynomials in $\xi_\mathrm{top}$ and $\xi_\mathrm{ext}$ and the coefficients are computed with a tensor Gauss-Hermite quadrature. With $N+1$ Gauss nodes per germ and an order-$N$ expansion, the moments in Eqs.~\eqref{eq:pce_mean}-\eqref{eq:pce_var} coincide with the quadrature moments, i.e. the model is evaluated at the $(N+1)^2$ combinations of the parameter values $\exp(\mu_{\ln,j}+\sigma_{\ln,j}\xi_{j,i})$ and the predictive mean and variance are the weighted mean and variance of these responses. An order-4 rule with $5\times5=25$ nodes is used.A convergence study for the required degree of the PCE is given in Appendix~\ref{ap:pce_order}.}
	
	The total predictive variance at sensor $k$ and time $t$ then follows directly from the decomposition in Eq.~\eqref{eq:variance decomposition}, with the \correction{embedded contribution} now attributed to MFU:
	\begin{equation}
		\Sigma_\mathcal{M}(t,k;\boldsymbol{\theta}_\mathrm{ext}) = \sigma^2_\mathrm{PCE}(t,k;\boldsymbol{\theta}_\mathrm{ext}) + \sigma_k^2.
	\end{equation}
	\review{We will assume additive \correction{residual} noise to be normally distributed. Additionally, in the absence of knowledge about the \correction{push-forward} predictive distribution, the Gaussian is the maximum-entropy distribution for fixed mean and variance, and thus introduces the least additional structure \citep{Jaynes1957}. \correction{The push-forward of the lognormal factors is not necessarily Gaussian. It is retained because the residual uncertainty is itself modelled as Gaussian noise, and the moment-matched Gaussian is therefore used both in the likelihood and in all validation metrics.} Under these assumptions, the log-likelihood replacing Eq.~\eqref{eq:log_likelihood} follows the independent normal likelihood of \citep{AndresArcones2026},}
	\begin{equation}
		\ell\left(\boldsymbol{\theta}_\mathrm{ext}, \left\lbrace\sigma_k^2\right\rbrace \mid \mathcal{Y}\right)
		= -\frac{1}{2}\sum_{k=1}^{N_k}\sum_{t=1}^{N_t}\left[\ln\left(2\pi\,\Sigma_\mathcal{M}(t,k;\boldsymbol{\theta}_\mathrm{ext})\right)
		+ \frac{\left[y(t,k) - m_\mathrm{PCE}(t,k;\boldsymbol{\theta}_\mathrm{ext})\right]^2}{\Sigma_\mathcal{M}(t,k;\boldsymbol{\theta}_\mathrm{ext})}\right].
		\label{eq:log_likelihood_mfu}
	\end{equation}
	The corresponding MLE problem is
	\begin{equation}
		\left(\hat{\boldsymbol{\theta}}_{\mathrm{ext},\mathrm{MLE}},\, \{\hat{\sigma}_{k,\mathrm{MLE}}^2\}\right)
		= \underset{(\boldsymbol{\theta}_\mathrm{ext},\,\left\lbrace\sigma_k^2\right\rbrace)\,\in\,\Theta_\mathrm{ext}\times(\mathbb{R}^+)^{N_k}}{\arg\max}
		\;\ell\left(\boldsymbol{\theta}_\mathrm{ext}, \left\lbrace\sigma_k^2\right\rbrace \mid \mathcal{Y}\right),
		\label{eq:mle_problem_mfu}
	\end{equation}
	where $\boldsymbol{\theta}_\mathrm{ext}$ enters the objective through both $m_\mathrm{PCE}$ and $\Sigma_\mathcal{M}$, since the PCE variance constitutes an additional model-dependent contribution to the total uncertainty. The PCE is reconstructed at each function evaluation of the optimizer using $Q$ quadrature nodes associated with $\pi_{\tilde{\boldsymbol{\theta}}}(\boldsymbol{\theta}_\delta)$, and the optimization is carried out with the Nelder-Mead algorithm \citep{Nelder1965}. \review{If the structure of the \correction{predictive distribution} is known a priori, other likelihood models can be derived analogously.}
	
\begin{figure}[!htbp]
		\centering
		\includegraphics[width=\textwidth]{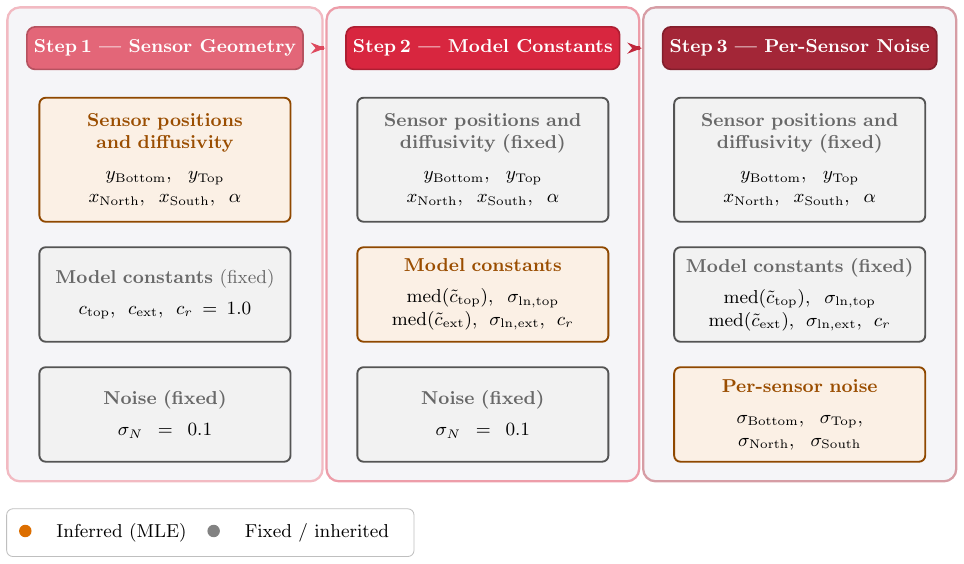}
		\caption{Flowchart for the proposed 3-steps parameter calibration framework}\label{fig:3steps_parameter_calibration}
	\end{figure}

	\paragraph{Three-step parameter calibration}
	\review{The thermal model presents three broad sources of uncertainty: geometric uncertainty resulting from simplifications in the model representation, uncertainty associated with the physical parameters governing the model response, and residual variability arising from modelling assumptions and unaccounted effects. The proposed three-step calibration framework addresses these sources in a sequential manner to reduce parameter interactions and improve the identifiability of the different uncertainty contributions. In the present application, geometric uncertainty is represented through both geometric variables (e.g., sensor positions) and \correction{the correction factors $c_\mathrm{top}$, $c_\mathrm{ext}$ and $c_r$}, rather than being restricted to a single source.} Figure~\ref{fig:3steps_parameter_calibration} provides a schematic overview.
	
	In the first step, the free coordinate of the sensor positions is calibrated. As established in Section~\ref{sec:sensitivity}, the sensor position is the dominant driver of model predictions. Discrepancies between the sensor locations reported in the bridge documentation and those of the virtual sensors in the FEM model must therefore be resolved before any other parameter can be reliably identified, as errors in position propagate into all subsequent estimates. Additionally, the thermal diffusivity parameter $\alpha$ is also included for calibration. The remaining constants \correction{$c_\mathrm{top}$, $c_\mathrm{ext}$} and $c_r$ are set to their nominal value 1.0 (implying no correction) and the noise is treated as isotropic and homoscedastic with $\sigma_N = 0.1$~°C, consistent with the sensor manufacturer specifications. The absence of stochastic extensions allows this step to be cast as the standard MLE problem of Eq.~\eqref{eq:mle_problem}, with the log-likelihood of Eq.~\eqref{eq:log_likelihood}.
	
	After the first step, the inferred sensor coordinates and diffusivity values are fixed for the remaining of the calibration. In the second step, a subset of the parameters of interest $\boldsymbol{\theta}$ are augmented with the embedded stochastic extension $\boldsymbol{\theta}_\delta$, and the MFU-aware MLE problem of Eq.~\eqref{eq:mle_problem_mfu}, with log-likelihood from Eq.~\eqref{eq:log_likelihood_mfu}, is solved\correction{, with the noise kept at the nominal value $\sigma_N=0.1$~°C of the first step}. This step yields the parameter configuration that best represents the observations within the limits of what the model can structurally reproduce. The model-form uncertainty is encoded in the variance induced by $\boldsymbol{\theta}_\delta$ and can be propagated accordingly. Nevertheless, due to the inherent discrepancy between the model response and the observations, the resulting predictive distribution may not fully account for the observed variability.
	
	The third step addresses this residual by inferring the per-sensor noise levels $\left\lbrace\sigma_k^2\right\rbrace$, with all other parameters fixed at their previously calibrated values. The MFU-aware formulations are retained, since the stochastic extensions from the second step are preserved. Because the model response does not change with variations in $\left\lbrace\sigma_k^2\right\rbrace$ alone, this step carries the lowest computational cost of the three. It quantifies the variance that the model cannot explain through its parameters or structure, capturing instead the sensor- and data-specific residual. Unlike the model-form uncertainty encoded in $\boldsymbol{\theta}_\delta$, this residual cannot be straightforwardly propagated to unseen conditions, as it is tied to the specific sensors and observations used for training. \correction{The noise levels are bounded from below by the measurement uncertainty of the sensors, $\sigma_k\geq\sigma_\mathrm{meas}=0.1$~°C, so that $\sigma_k^2=\sigma_\mathrm{meas}^2+\sigma_{\mathrm{res},k}^2$ with a residual model error $\sigma_{\mathrm{res},k}^2\geq0$.} %

	\correction{Two baselines are used to separate the contributions of the method. The \textit{deterministic-correction} baseline follows the same three calibration steps, with the same parameters and data, but estimates $c_\mathrm{top}$, $c_\mathrm{ext}$ and $c_r$ as deterministic values in the second step ($\sigma_{\ln,\mathrm{top}}=\sigma_{\ln,\mathrm{ext}}=0$). It therefore differs from the embedded model only in that the uncertainty of the factors is not embedded in the model, which isolates the effect of the embedding from that of estimating the factors. The \textit{only sensors} baseline fixes $c_\mathrm{top}=c_\mathrm{ext}=c_r=1$ after the first step and calibrates only the noise. All model discrepancy is then absorbed by the observational noise, which corresponds to the classical treatment.}
	
	An alternative framework in which all parameters are inferred simultaneously could be considered. However, such an approach would compromise the identifiability \citep{Arendt2012}, falsifiability \citep{Cademartori2023}, and parsimony \citep{Gori2024} of the identification procedure, particularly in the presence of substantial model-data discrepancies. First, jointly calibrating the sensor positions with the remaining parameters would cause the inference to be dominated by the sensor coordinates, as the sensitivity analysis shows that they exert the largest influence on the model response. Geometric adjustments may compensate for discrepancies arising from other sources, such as physical parameter uncertainty or model-form error, making the attribution of the identified corrections less interpretable. Estimating the sensor locations in a preliminary step reduces this confounding effect and allows the subsequent calibration phases to estimate the remaining parameters conditional on the inferred geometry. Second, simultaneously estimating the observational noise and the embedded variance parameters leads to identifiability issues when large discrepancies are present. In such situations, the likelihood cannot reliably attribute variability to either observational noise or model-form uncertainty unless the heteroscedastic structure induced by the propagated MFU is sufficiently pronounced in the model response.
	
	\subsection{Results}
	\label{sub:calibration_results}
	\correction{The parameters are estimated sequentially by maximizing the log-likelihood of Equation~\eqref{eq:log_likelihood} in the first step and the MFU-aware log-likelihood of Equation~\eqref{eq:log_likelihood_mfu} in the second and third steps, with the Nelder-Mead algorithm of the \texttt{minimize} routine of \texttt{scipy}, a tolerance of $10^{-4}$ in the parameter values or in the log-likelihood, and at most 700 function evaluations in the first step and 1500 in the second and third steps.}

	In the first phase, the sensor positions and the diffusivity $\alpha$ are inferred assuming $\sigma_N = 0.1$~°C. \correction{The free coordinate of each sensor is its depth measured from the inner surface, which spans the full width of the wall for $T_\mathrm{South}$, $T_\mathrm{North}$ and $T_\mathrm{Top}$, and the equivalent range plus the distance to the inner wall for $T_\mathrm{Bottom}$, so that the sensors remain within the virtual structure while covering the geometric uncertainties (Table~\ref{tab:inference_comparison_no_hr}).} \correction{The diffusivity $\alpha$ is pushed to the lower limit of $0.8$~mm$^2$/s to compensate for the discrepancy at the bottom sensor, a clear indication of unavoidable MFU, which is also partially explained by the limited identifiability of diffusivity and sensor depth, since both control the attenuation and phase shift of the temperature signals. Its value is therefore not representative and should not be trusted for other predictions, although its influence in the present configuration is minimal as per the sensitivity analysis. Likewise, the depth of $T_\mathrm{South}$ converges to the lower bound of $3.45$~m, reflecting modelling simplifications such as the absence of radiation on the sides of the bridge and wind dynamics that may favour convection on the southern side. These errors are absorbed by the uncertainty estimates of the subsequent steps, in which the estimated coordinates are the virtual sensor positions.} \correction{In the classical extension, sensor-specific noise levels $\sigma_k$ are then inferred under the assumption of homoscedastic Gaussian noise, so that all unexplained discrepancies are attributed to the noise, which absorbs measurement noise, model discrepancy and unmodelled dynamics alike. These values (``Only sensors'' in Table~\ref{tab:inference_comparison_no_hr}) are the baseline for the following sections.}
	
	\begin{table}[htbp]
\centering
\caption{Calibration results per step. The embedded model has two independent lognormal factors, $\tilde c_\mathrm{top}$ of the convective exchange at the deck top and $\tilde c_\mathrm{ext}$ of all other external faces; each is described by its median $\mathrm{med}(\tilde c)$ and the standard deviation $\sigma_{\ln}$ of its logarithm, and its mean in linear space is a derived value. Step~1 is shared by all models. The noise is bounded from below by the measurement uncertainty of the sensors, 0.1~$^\circ$C. The model without corrections was calibrated before the bound was introduced; its noise values are all above 0.1~$^\circ$C.}
\label{tab:inference_comparison_no_hr}
\begin{tabular}{lcccc}
\hline
\textbf{Variable} & \textbf{Domain} & \textbf{Only sensors} & \textbf{Det. corrections} & \textbf{Embedded} \\
\hline
\multicolumn{5}{l}{\textit{Step 1: sensor positions and diffusivity}} \\
\hline
$\alpha$ [m$^2$/s] & (8e-7, 2e-6) & \multicolumn{3}{c}{8.00e-07} \\
Sensor $T_\mathrm{Bottom}$ - y [m] & (-5.6, -5.1) & \multicolumn{3}{c}{-5.264} \\
Sensor $T_\mathrm{South}$ - x [m] & (3.45, 3.8) & \multicolumn{3}{c}{3.450} \\
Sensor $T_\mathrm{North}$ - x [m] & (1.9, 2.15) & \multicolumn{3}{c}{2.081} \\
Sensor $T_\mathrm{Top}$ - y [m] & (-0.3, -0.05) & \multicolumn{3}{c}{-0.171} \\
\hline
\multicolumn{5}{l}{\textit{Step 2: convection and shortwave constants}} \\
\hline
$c_\mathrm{top}$ / $\mathrm{med}(\tilde c_\mathrm{top})$ & (0.01, 4.0) & 1 (fixed) & 0.381 & 0.447 \\
$\sigma_{\ln,\mathrm{top}}$ & (0, 5) & -- & -- & 0.468 \\
\quad mean of $\tilde c_\mathrm{top}$ (derived) & -- & -- & -- & 0.50 \\
$c_\mathrm{ext}$ / $\mathrm{med}(\tilde c_\mathrm{ext})$ & (0.01, 4.0) & 1 (fixed) & 0.874 & 1.672 \\
$\sigma_{\ln,\mathrm{ext}}$ & (0, 5) & -- & -- & 2.258 \\
\quad mean of $\tilde c_\mathrm{ext}$ (derived) & -- & -- & -- & 21.38 \\
$c_r$ & (0.01, 4.0) & 1 (fixed) & 0.688 & 0.706 \\
\hline
\multicolumn{5}{l}{\textit{Step 3: sensor noise}} \\
\hline
$\sigma_{\mathrm{Bottom}}$ [$^\circ$C] & (0.1, 10.0) & 0.175 & 0.164 & 0.139 \\
$\sigma_{\mathrm{Top}}$ [$^\circ$C] & (0.1, 10.0) & 1.249 & 0.821 & 0.284 \\
$\sigma_{\mathrm{North}}$ [$^\circ$C] & (0.1, 10.0) & 0.264 & 0.255 & 0.227 \\
$\sigma_{\mathrm{South}}$ [$^\circ$C] & (0.1, 10.0) & 0.444 & 0.449 & 0.380 \\
\hline
\end{tabular}
\end{table}

\begin{figure}[!htbp]
		\centering
		\includegraphics[width=\textwidth]{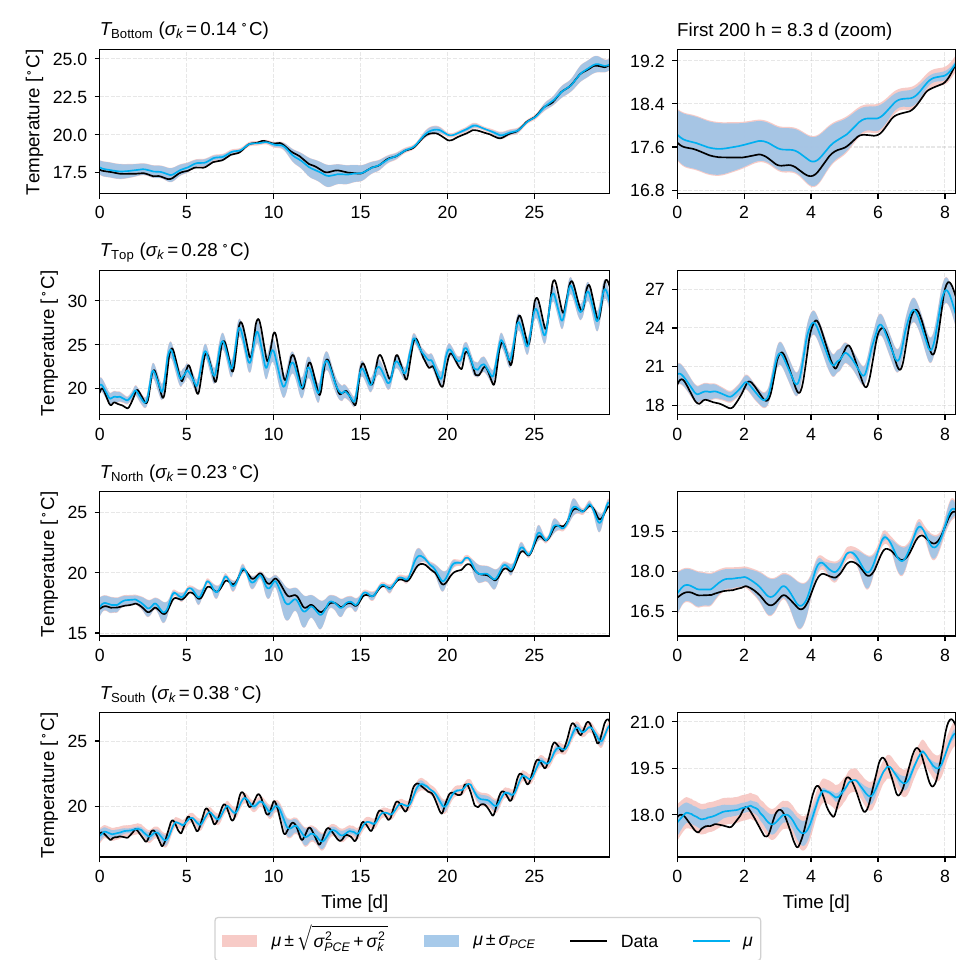}
		\caption{\correction{Calibrated predictive distribution} for June~2024 after the full calibration pipeline\correction{ (embedded model, calibrated noise). Left: full month; right: first 200~h (8.3~days). From top to bottom: $T_\mathrm{Bottom}$, $T_\mathrm{Top}$, $T_\mathrm{North}$, $T_\mathrm{South}$}}\label{fig:inference_predictions}
	\end{figure}

	In the second phase, the \correction{convection factors $c_\mathrm{top}$ and $c_\mathrm{ext}$ are} selected for stochastic embedding because \correction{they act} directly on the surface boundary conditions where the dominant modelling discrepancies originate and are proven to produce a sensitive response. \review{The errors committed by fixing the sensor positions and $\alpha$ to the values of the first phase are considered part of the MFU.} \correction{Each embedded factor is an independent lognormal random variable, $\ln\tilde c_j\sim\mathcal{N}(\mu_{\ln,j},\sigma_{\ln,j}^2)$, $j\in\lbrace\mathrm{top},\mathrm{ext}\rbrace$, parametrized in log space by its median $\mathrm{med}(\tilde c_j)=\exp(\mu_{\ln,j})$, which takes the place of $c_j$ and is restricted to the domain $(0.01,4)$ of the deterministic factor, and by the standard deviation $\sigma_{\ln,j}\in(0,5)$ of its logarithm. The mean of $\tilde c_j$ in linear space follows as $\exp(\mu_{\ln,j}+\sigma_{\ln,j}^2/2)$ and is reported as a derived value. The shortwave factor $c_r$ is estimated deterministically in the same step.}%
Environmental measurements from June~2024 and their corresponding sensor observations are used as the training dataset, denoted $\mathcal{Y}_{\text{train}}$. The variance introduced through this stochastic extension is interpreted as \correction{embedded model error}, since it is embedded directly within the model while the prescribed sensor noise is assumed to remain correct. This interpretation does not imply that the discrepancies between predictions and observations are necessarily caused by a miscalibration of these parameters. Rather, variability in these parameters is used as a proxy representation of model-form uncertainty. Subsequently, the sensor noise levels are inferred in the third step while keeping the previously calibrated constants fixed. Table~\ref{tab:inference_comparison_no_hr} reports the inferred parameters at each step, and Figure~\ref{fig:inference_predictions} shows the \correction{calibrated predictive distribution} after the complete calibration process.
	
	\correction{The two embedded factors differ clearly (Table~\ref{tab:inference_comparison_no_hr}). The top factor $c_\mathrm{top}$ halves the convective exchange, which is compatible with an additional thermal resistance of the pavement and asphalt layers that are not modelled. The discrepancy at the deck top is thus mainly a systematic reduction of the exchange. The factor of the other external faces has a larger median than $c_\mathrm{ext}$ of the baseline and a wide spread, which indicates a discrepancy there that varies in time, most likely the lack of radiation at the southern sensor. The shortwave factor $c_r<1$, almost unchanged by the embedding, shows that the model overestimates the solar heat input at the deck top.Since part of the model discrepancy is absorbed by $c_r$ and by the log-standard deviations $\sigma_{\ln,\mathrm{top}}$ and $\sigma_{\ln,\mathrm{ext}}$ of the correction factors, these quantities express all the discrepancies present during calibration, not only the uncertainty of the convective exchange. They should therefore be interpreted as effective uncertainties rather than as physical uncertainties of the convective coefficients.}

	\correction{The distribution of $\tilde c_\mathrm{ext}$ is wide, leading to implausible values for a convection coefficient. The width is possible because the surface temperature saturates between an insulated surface for $\tilde c_\mathrm{ext}\to0$ and a surface at air temperature for $\tilde c_\mathrm{ext}\to\infty$, so that a wide distribution of $\tilde c_\mathrm{ext}$ produces a bounded predictive spread that follows the difference between these two regimes over the day. $\tilde c_\mathrm{ext}$ is therefore a carrier of model discrepancy, not an uncertain convection coefficient. The embedding has its largest effect on the noise of $T_\mathrm{Top}$. The calibrated noise variance is 89\% lower than in the deterministic-correction baseline. The classical treatment attributes this discrepancy entirely to observational noise, whereas the embedding shifts part of it to the uncertainty of the correction factors. The misfit is thus moved into the embedded predictive variance. Its benefit is that this variance is not constant: it varies in time with the sensitivity of the model output to the factors, it is carried forward over the time steps by the model dynamics, and it can be propagated to other quantities of interest. The noise of $T_\mathrm{Bottom}$ and $T_\mathrm{North}$ decreases slightly, because the uncertainty of the external-face factors also contributes to the predictive variance at these sensors, so that less of their residual has to be attributed to noise. $T_\mathrm{South}$ retains the largest noise relative to its predictive variance. Model improvements should therefore focus on the southern side of the cross-section, for example on radiation and sun-shade dynamics.}
	
	\subsection{Discussion on model prediction and parameter interpretation}
	An important practical question concerns how predictions should be generated using models calibrated in this way. In the proposed framework, predictive quantities are obtained by propagating the \correction{calibrated distribution} of the embedded parameters through the model. The interpretation of the resulting predictive uncertainty depends on the role assigned to the stochastic embedding. If the embedding is regarded as an intrinsic component of the model, the uncertainty associated with realizations of the embedded stochastic process may be interpreted as aleatoric, while uncertainty in its inferred hyperparameters remains epistemic. Alternatively, if the embedded parameters are regarded as uncertain parameters representing MFU, the uncertainty arising from their \correction{calibrated} distribution is interpreted as epistemic. The distinction between epistemic and aleatoric uncertainty is not unique, but depends on the adopted modelling interpretation \citep{Kiureghian2009}. \correction{Neither label describes the embedded variance of this work without qualification. It does not arise from parameter inference, since all parameters, including the spreads $\sigma_{\ln,\mathrm{top}}$ and $\sigma_{\ln,\mathrm{ext}}$, are point estimates and their estimation uncertainty is much smaller, and it is not irreducible, since it would change with an improved model. The Laplace standard errors of the spreads, from the inverse of the negative Hessian of the log-likelihood at the optimum, are 0.012 for $\sigma_{\ln,\mathrm{top}}=0.47$ and 0.036 for $\sigma_{\ln,\mathrm{ext}}=2.26$. Inflated by $\sqrt{\tau}\approx4$--$12$ for the autocorrelation of the residuals (Section~\ref{sub:bayesian_calibration}), they remain at least 3.4 and 5.2 times smaller than the spreads. Calibrations on training windows of 1 week to 2 months, starting between 31 May and 21 June 2024, give $\sigma_{\ln,\mathrm{top}}$ between 0.32 and 0.56 and $\sigma_{\ln,\mathrm{ext}}$ between 1.8 and 2.4, except for one window in which $\sigma_{\ln,\mathrm{ext}}$ reaches the upper bound of its domain; neither spread approaches zero in any window. It is therefore referred to as embedded model error, and the predictive distribution combines it with the residual model error and the measurement noise, both contained in $\sigma_k^2$.}
	
	When predictions concern quantities that are directly derived from the calibrated model, such as temperatures at new times or at locations governed by the same physical model, the propagated \correction{embedded distribution} provides an estimate of the uncertainty associated with applying the calibrated model beyond the observed data. Propagating the full predictive distribution therefore provides a principled estimate of predictive reliability. However, this interpretation remains conditional on the adopted model structure. The embedded parameters characterize the component of model-form uncertainty represented by the chosen stochastic embedding, but they cannot account for all possible structural deficiencies or modelling assumptions outside this representation. Consequently, the predictive uncertainty obtained through MFU-aware calibration should be interpreted as an estimate of the uncertainty induced by the adopted model class rather than as a comprehensive bound on all modelling errors.
	
	A related issue concerns the interpretation and transferability of the calibrated parameters. The parameters obtained through calibration should be interpreted within the context of the adopted model formulation rather than as universal material or structural properties. When model-form discrepancies are present, calibrated parameters may compensate for deficiencies in the model representation \citep{Kleijn2012}. Parameters identified using one modelling approach may contain a bias associated with the specific assumptions and simplifications of that model, which can limit their direct transfer to alternative formulations. From this perspective, calibration aims to identify parameter distributions that provide reliable predictions within the adopted modelling framework, rather than to determine model-independent parameter values.

	\section{Variance decomposition, uncertainty attribution and model validation} 
	The hierarchical structure adopted in the calibration stage provides a natural and rigorous framework for quantifying the different sources of predictive uncertainty. In particular, a decomposition of the predictive variance enables the attribution of the total uncertainty to its constituent components, allowing us to assess to what extent the discrepancy between model predictions and observations can be explained by \correction{the embedded model error, as opposed to residual model error} or measurement noise. This distinction is key for the proper assessment of uncertainty in model predictions \citep{Oliver2015}. We complement these analysis with a direct evaluation of the Kolmogorov-Smirnov deviation \citep{Villani2026}, which offers a quantitative measurement of the remaining discrepancy in the calibrated model and a non-parametric validation of the predictive distributions over the annual cycle.
	
	\subsection{Variance decomposition}
	\label{sec:variance_decomposition}
	\correction{Uncertainty in predictive models is often classified into aleatoric and epistemic contributions \citep{Walker2003}. The former is generally associated with intrinsic variability in the observations, whereas the latter reflects incomplete knowledge of the system, for instance due to uncertain model parameters or model-form discrepancy. However, these categories are not directly adopted in the present framework, since the computed uncertainty components do not correspond unambiguously to this conventional distinction. In particular, the embedded variance represents neither parameter-estimation uncertainty nor irreducible variability. Instead, the proposed framework distinguishes between embedded model error, residual model error and measurement noise. The stochastic embedding introduced during calibration induces variability in the model response through the calibrated distribution of the embedded parameters, thereby representing the embedded model error. Residual model error and measurement noise are represented separately through the additive sensor noise model. The resulting predictive variance decomposition can therefore be derived analytically from the probabilistic structure of the model, without requiring replicated observations.}
	
	\subsubsection{Formulation}
	Let $\tilde{y}_\mathcal{M}(t,k)$ denote the predictive temperature at sensor $k$ and time $t$ obtained from model $\mathcal{M}$ \correction{with the calibrated parameters $\hat{\boldsymbol{\theta}}_\mathrm{ext}$ and the embedded random parameters $\tilde{\boldsymbol{\theta}}\sim\pi(\cdot|\hat{\boldsymbol{\theta}}_\mathrm{ext})$}. The predictive model is assumed to follow an additive structure, 
	\begin{equation}
		\tilde{y}_\mathcal{M}(t,k)=\tilde{T}_\mathcal{M}(t,k;\tilde{\boldsymbol{\theta}})+\varepsilon_k, 
		\qquad 
		\varepsilon_k \sim \mathcal{N}(0,\sigma_k^2),
	\end{equation}
	where $\tilde{T}_\mathcal{M}(t,k;\tilde{\boldsymbol{\theta}})$ represents the stochastic model response induced by \correction{the embedded parameters}, and $\varepsilon_k$ is a sensor-specific noise term independent of $\tilde{\boldsymbol{\theta}}$. The \correction{predictive} variance is defined as
	\begin{equation}
		\Sigma_\mathcal{M}(t,k):=\mathrm{Var}\left(\tilde{y}_\mathcal{M}(t,k)\right).
	\end{equation}
	Applying the law of total variance \citep{Blitzstein2019} with respect to the \correction{distribution} of $\tilde{\boldsymbol{\theta}}$ yields
	\begin{equation}
		\mathrm{Var}(\tilde{y}_\mathcal{M}(t,k))=
		\underbrace{\mathbb{E}_{\tilde{\boldsymbol{\theta}}}\left[\mathrm{Var}\left(\tilde{y}_\mathcal{M}(t,k)\mid\tilde{\boldsymbol{\theta}}\right)\right]}_{\correction{\text{Residual and measurement variance}}}
		+
		\underbrace{\mathrm{Var}_{\tilde{\boldsymbol{\theta}}}\left[\mathbb{E}\left(\tilde{y}_\mathcal{M}(t,k)\mid\tilde{\boldsymbol{\theta}}\right)\right]
		}_{\correction{\text{Embedded variance}}}.
	\end{equation}
	The \correction{first term} quantifies the intrinsic variability of the observations conditional on fixed model parameters. It captures \correction{the} measurement noise and \correction{the residual model error, i.e. the} disturbances not explicitly represented in the model\correction{ or in the embedding}. In contrast, the second term represents the variability of the predictive mean associated with the embedded parameters $\tilde{\boldsymbol{\theta}}$. It is the part of the model discrepancy that is represented by the embedding, and it does not vanish with more data of the same kind, it can be reduced by improving the model. The embedding parameters are maximum likelihood estimates $\hat{\boldsymbol{\theta}}_\mathrm{ext}$, so the estimation uncertainty of these parameters is not part of this term and it is neglected here. Under the adopted additive formulation, the conditional moments satisfy
	\begin{equation}
		\mathrm{Var}\left(\tilde{y}_\mathcal{M}(t,k)\mid\tilde{\boldsymbol{\theta}}\right)=\sigma_k^2,
		\qquad
		\mathbb{E}\left(\tilde{y}_\mathcal{M}(t,k)\mid\tilde{\boldsymbol{\theta}}\right)
		=\tilde{T}_\mathcal{M}(t,k;\tilde{\boldsymbol{\theta}}),
	\end{equation}
	so that the predictive variance admits the explicit decomposition
	\begin{equation}
		\mathrm{Var}(\tilde{y}_\mathcal{M}(t,k))=
		\sigma_k^2
		+
		\mathrm{Var}\left(\tilde{T}_\mathcal{M}(t,k;\tilde{\boldsymbol{\theta}})\right)=\underbrace{\sigma_k^2}_{\correction{\text{Residual and measurement variance}}}
	+\underbrace{\sigma^2_\mathrm{PCE}(t,k;\correction{\hat{\boldsymbol{\theta}}_\mathrm{ext}})}_{\correction{\text{Embedded variance}}}.
		\label{eq:variance decomposition}
	\end{equation}
	Equivalently,
	\begin{equation}
		\Sigma_\mathcal{M}(t,k)=\sigma^2_\mathrm{PCE}(t,k;\correction{\hat{\boldsymbol{\theta}}_\mathrm{ext}})+\sigma_k^2.
	\end{equation}
	\correction{With the lower bound of the noise at the measurement uncertainty of the sensors (Section~\ref{sub:bayesian_calibration}), the noise variance is split further into measurement noise and residual model error, $\sigma_k^2=\sigma^2_\mathrm{meas}+\sigma^2_{\mathrm{res},k}$ with $\sigma_\mathrm{meas}=0.1$~°C, which gives the three-way decomposition
	\begin{equation}
		\Sigma_\mathcal{M}(t,k)=\underbrace{\sigma^2_\mathrm{PCE}(t,k;\hat{\boldsymbol{\theta}}_\mathrm{ext})}_{\text{Embedded model error}}+\underbrace{\sigma^2_{\mathrm{res},k}}_{\text{Residual model error}}+\underbrace{\sigma^2_\mathrm{meas}}_{\text{Measurement noise}}.
		\label{eq:three_way_split}
	\end{equation}
	For the models without embedding, $\sigma^2_\mathrm{PCE}=0$ and the whole discrepancy is contained in $\sigma^2_{\mathrm{res},k}$. The value $\sigma_\mathrm{meas}=0.1$~°C is the upper bound of the measurement uncertainty given by the manufacturer, and the noise of the 30-min values obtained from the 10-Hz signal is likely smaller. The split between residual model error and measurement noise is therefore conditional on this value, whereas the embedded share is not.}
	
	We define the ratio between \correction{the noise variance (residual model error and measurement noise)} and total predictive variance as
	\begin{equation}
		S_{\text{noise},k}=\frac{\sigma_k^2}{\sigma^2_\mathrm{PCE}(t,k;\correction{\hat{\boldsymbol{\theta}}_\mathrm{ext}})+\sigma_k^2}
		=
		\frac{\sigma_k^2}{\Sigma_\mathcal{M}(t,k)}.
	\end{equation}
	A large value of $S_{\text{noise},k}$ indicates that a substantial portion of the predictive variance is attributed to the additive noise term rather than to \correction{the embedded} variability. This suggests that the model is unable to explain the observed discrepancies and variability exclusively through the stochastic embedding, and that the additional noise absorbs effects that remain structurally unmodelled or uncontrolled. The index can therefore be interpreted as the relative magnitude of the residual noise compared with the uncertainty propagated from the embedded parameters. \correction{Its complement $S_{\mathrm{emb},k}=1-S_{\text{noise},k}$ is the embedded share of the predictive variance.}
	
	Unless otherwise specified, the variance components and the index $S_{\text{noise},k}$ are reported as their median over the time window of interest. The median provides a robust summary measure that reduces the influence of transient local effects and extreme events.
	
	\subsubsection{Results and discussion}
\begin{figure}[!htbp]
		\centering
		\includegraphics[width=\textwidth]{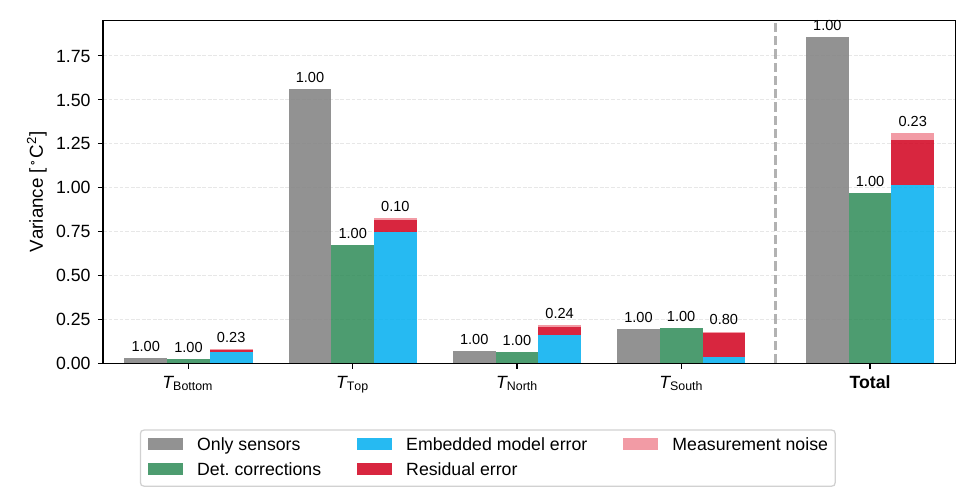}
		\caption{Median variance decomposition after calibrating the noise component for the \correction{models} without embedding (\correction{in green, only sensors and deterministic corrections, noise variance only}) and with embedding (decomposed in \correction{embedded model error in blue, residual model error in dark red and measurement noise in light red, per Equation~\ref{eq:three_way_split}}). Median values across the month of June~2024. The \correction{noise share} $S_{noise}$ is indicated above their respective bars} \label{fig:variance_decomposition_histogram}
	\end{figure}

	\correction{Figure~\ref{fig:variance_decomposition_histogram} shows the variance decomposition at each sensor for the model calibrated with data from June~2024, together with the predictive variance of the two baselines without embedding, which is their calibrated noise (Table~\ref{tab:inference_comparison_no_hr}).} As stated in Equation~\ref{eq:variance decomposition}, the \correction{embedded} variance arises exclusively from the push-forward uncertainty of the extended model parameters, whereas the \correction{residual and measurement} variance is determined entirely by the homoscedastic Gaussian noise contribution\correction{, which is split according to Eq.~\eqref{eq:three_way_split}}.
	
	\correction{At $T_\mathrm{Top}$, where most of the predictive uncertainty is concentrated, the predictive variance of the embedded model lies between those of the two baselines, at $T_\mathrm{Bottom}$ and $T_\mathrm{North}$ it is larger, and at $T_\mathrm{South}$ comparable. Except at $T_\mathrm{South}$, most of it is embedded variance, which is heteroscedastic and can be propagated to the QoIs. At $T_\mathrm{Top}$, the embedding accounts for most of the discrepancy, and, at $T_\mathrm{Bottom}$ and $T_\mathrm{North}$, it represents discrepancies beyond the deck top through the factor of the external faces. At $T_\mathrm{South}$, the residual model error dominates, owing to the remaining model discrepancies discussed above. It is remarkable that calibrating the noise components without the embedded model error but after estimating the deterministic corrections tends to provide smaller variances than with the embedding. This is in big part due to the heteroscedastic nature of the propagated predictive variance with embedding, which may lead to overcompensate parts of the prediction which are not well represented by the model. However, it must be restated that the variance estimated without the embedding cannot be propagated in general. Although the variance decomposition is not a complete analysis of the uncertainties of the calibrated model, it attributes the predictive variance to its sources and informs the assessment of their relevance.}
	
	\subsection{Kolmogorov-Smirnov discrepancy analysis}
	\label{sec:ks_discrepancy}
	Classical error metrics such as RMSE ignore the predictive variance entirely, reducing model quality to a measure of mean bias, while variance-informed metrics such as the Mahalanobis distance are often difficult to interpret under model-form uncertainty, which is absorbed into the fitted variance and masks systematic misfits at the sensor level. \citet{Villani2026} introduce a discrepancy metric based on the Kolmogorov-Smirnov (KS) test to assess whether a stochastic model prediction can account for observed sensor readings. Comparing the discrepancy of predictions with nominal and with calibrated per-sensor noise across different times of the year helps identify when the model can be trusted and informs potential improvements.
	\subsubsection{Formulation}
	Let us consider the empirical cumulative distribution function (CDF) of the squares of the normalized residuals $\sigma(t,k)^{-1}\left(y(t,k) - \tilde T_\mathcal{M}(t,k)\right)$ at time index $t$ and sensor $k$ considering model $\mathcal{M}$ as 
	\begin{equation}
		\hat F_{\mathcal{M}}(\eta,k) := \frac{1}{n} \left| \big\{ i\in[1,n]\cap \mathbb{N} \mid (y(t,k)-\tilde T_\mathcal{M}(t,k))^2 < \eta\sigma(t,k)^2\big\}\right|,
	\end{equation}
	where the dependency on the calibrated parameters $\bm{\theta}$ will be \review{omitted} to simplify notation. For a model $\mathcal{M}$ that replicates the real system's response $T(t,k)$ perfectly, $\hat F_{\mathcal{M}}(\eta,k)$ coincides with the CDF $F_{\chi^2_1}$ of the normally distributed errors $\varepsilon(t,k)$, which is distributed as a $\chi^2$ distribution of one degree of freedom. The two-sided KS test considers the statistic
	\begin{equation}
		D_{\mathcal{M},k} := \sup_{\eta\in\mathbb{R}_+} |F_{\chi^2_1}(\eta)-\hat F_{\mathcal{M}}(\eta,k)|,
	\end{equation}
	which for $n\to\infty$ follows a Kolmogorov distribution $K$ with CDF $F_K$ such that $\sqrt{n}D\sim K$. This two-sided test is sensitive to any discrepancy in the \correction{predictive} distribution, either underestimation or overestimation of the variance in the predictions or errors in the mean values. A one-sided version that does not penalize larger variances can be built from the statistic
	\begin{equation}
		\bar{D}^{-}_{\mathcal{M},k} := \sup_{\eta\in\mathbb{R}_+} F_{\chi^2_1}(\eta)- \bar{F}_{\mathcal{M}}(\eta,k),
	\end{equation}
	where $\bar{F}_{\mathcal{M}}(\eta,k)$ considers only the upper bounds of the variance $\bar{\sigma}(t,k)^2>\sigma(t,k)^2$.
	
	The KS test rejects the null hypothesis $H_0$: the model $\mathcal{M}$ is indistinguishable from reality with a significance level $\correction{\gamma}\in]0,1[$ if $F_K(\sqrt{n}D_{\mathcal{M},k})>1-\correction{\gamma}$. Based on this, the \textit{deviation} is defined as the quantity $d$ required such that the KS test is passed as
	\begin{equation}
		d_{\mathcal{M},k} = \inf\left\{ \delta \in\mathopen]0,1\mathclose[ \Big| F_K\left(\sqrt{n} (D_{\mathcal{M},k}-\delta) \right) \leq 1-\correction{\gamma} \right\}.
	\end{equation}
	Values $d_{\mathcal{M},k}$ close to 0 indicate that the model $\mathcal{M}$ can plausibly reproduce the observations at sensor $k$, whereas $d_{\mathcal{M},k}\approx 1$ provides evidence against model adequacy. An analogous construction applies to the one-sided statistic $\bar{d}^{-}_{\mathcal{M},k}$ obtained from a one-sided KS test, in which only underdispersion of the predictive distribution is penalized, while overestimation of predictive uncertainty is intentionally disregarded.
	
	The deviation $d_{\mathcal{M},k}$ is defined for a fixed set of observations and corresponding predictive distributions. To assess the temporal evolution of model adequacy, we adopt a rolling-window formulation. For each sensor $k$, let $\{(y_i,t_i)\}_{i=1}^{N}$ denote the time-ordered observations. These data are partitioned into a sequence of overlapping windows indexed by $w\in\{1,\dots,W\}$, where each window contains $N_W$ consecutive observations and successive windows are shifted by a fixed increment $\Delta w$. For each window $w$, the deviation $d_{\mathcal{M},k}(w)$ is computed using only the observations and predictions restricted to that window.
	
	For models exhibiting substantial mismatch with the data ($d_{\mathcal{M},k}\approx 1$), the resulting sequence $\{d_{\mathcal{M},k}(w)\}_{w=1}^{W}$ may contain extreme values close to 1 that are sensitive to the particular realization of observations within individual windows. Although this sensitivity does not affect relative model comparisons conducted on identical datasets, it can hinder the identification of coherent temporal trends in model adequacy. To obtain a more robust characterization of the deviation evolution, we regularize the time series using rolling order statistics rather than a single-point smoothing operator. Specifically, let $L_{\mathrm{roll}}$ denote the number of consecutive deviation values included in each rolling estimate. This quantity is distinct from the observation-window size $N_W$: while $N_W$ determines the number of observations used to compute each individual deviation $d_{\mathcal{M},k}(w)$, $L_{\mathrm{roll}}$ specifies how many consecutive deviation values are aggregated to obtain the rolling statistics. Successive rolling estimates are evaluated by shifting the aggregation window by $\Delta w$.
	
	For each window index $w=1,\ldots,W-L_{\mathrm{roll}}+1$, we consider the subset
	\begin{equation}
	\left\{
	d_{\mathcal{M},k}(j):
	j=w,\ldots,w+L_{\mathrm{roll}}-1
	\right\},
	\end{equation}
	and define the rolling empirical quantiles
	\begin{equation}
		\widetilde{d}^{(q)}_{\mathcal{M},k}(w)
		=
		\operatorname{Quantile}_q
		\left\{
		d_{\mathcal{M},k}(j):
		j=w,\ldots,w+L_{\mathrm{roll}}-1
		\right\},
		\qquad
		q\in\{0.25,0.5,0.75\}.
	\end{equation}
	The case $q=0.5$ corresponds to the rolling median,
	\begin{equation}
		\widetilde{d}_{\mathcal{M},k}(w)
		=
		\widetilde{d}^{(0.5)}_{\mathcal{M},k}(w),
	\end{equation}
	which provides a robust measure of central tendency. The associated interquartile range, given by the $25^{\mathrm{th}}$ and $75^{\mathrm{th}}$ quantiles, complements the median by quantifying local variability in the deviation. Together, the triplet
	\begin{equation}
	\left(
	\widetilde{d}^{(0.25)}_{\mathcal{M},k},
	\widetilde{d}^{(0.5)}_{\mathcal{M},k},
	\widetilde{d}^{(0.75)}_{\mathcal{M},k}
	\right)
	\end{equation}
	provides a robust representation of both the central tendency and dispersion of the deviation across adjacent windows, reducing the influence of window-specific outliers while avoiding unnecessary loss of information. In the present study, the observation-window size was set to $N_W=30$ days, with successive windows shifted by $\Delta w=7$ days. The rolling quantiles were computed using $L_{\mathrm{roll}}=\correction{30}$ consecutive deviation values.
	
	\subsubsection{Results}
	\correction{Table~\ref{tab:deviation_trainingmonth} reports the KS-deviations for the training month (June~2024). With nominal noise, only the embedded model has moderate KS-deviations. Calibrating the residual variance of each sensor as noise lowers all  KS-deviations, because the Gaussian noise model can closely reproduce the empirical distribution of the residuals, but for the models without embedding this is achieved through large predictive variances that are entirely classified as noise. The one-sided deviations of the embedded model are close to zero at $T_\mathrm{Bottom}$ and $T_\mathrm{North}$, where its predictive distribution is over-dispersed.}

\begin{figure}[!htbp]
		\centering
		\includegraphics[width=\textwidth]{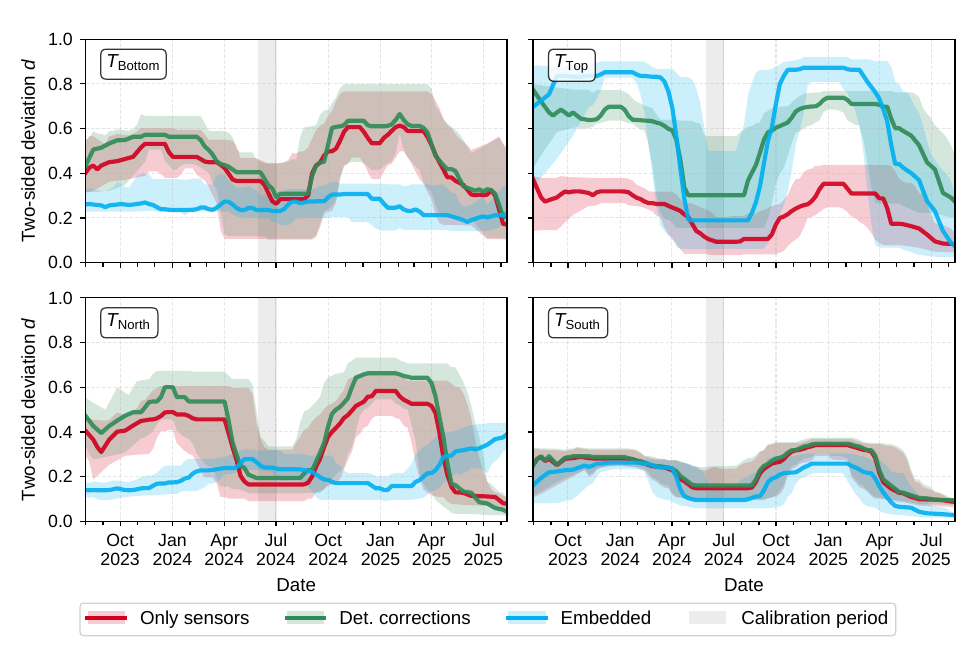}
		\caption{Temporal evolution of the rolling deviation quantiles $\left(\widetilde{d}^{(0.25)}_{\mathcal{M},k}, \widetilde{d}^{(0.5)}_{\mathcal{M},k}, \widetilde{d}^{(0.75)}_{\mathcal{M},k}\right)$ for each sensor from July 2023 to \correction{August 2025} with the parameters identified using the data of June~2024, corresponding to the two-sided KS metric. \correction{Calibrated noise for the three models. Lines: rolling median; shaded bands: interquartile range; the calibration period is marked}}\label{fig:deviation_timeline_comparison}
	\end{figure}

	\correction{Figure~\ref{fig:deviation_timeline_comparison} shows the rolling quantiles of the two-sided KS-deviation from July~2023 to August~2025 for the parameters calibrated with the data of June~2024 ($\Delta t=30$~min), which evaluates the temporal robustness of the calibrated models outside the calibration period.} \review{The two-sided diagnostic penalizes overly wide variances that do not fit the data sharply. \correction{Nevertheless, the homoscedastic noise calibrated for each sensor in the classical model produces a relatively broad predictive distribution that reduces the deviation at most sensors over the whole period.}} \correction{At $T_\mathrm{Bottom}$, $T_\mathrm{North}$ and $T_\mathrm{South}$, the embedded model has the lowest deviations of the three calibrations, and they change little between the seasons, whereas at $T_\mathrm{Top}$ its deviation is the largest, for reasons discussed in Section~\ref{sub:predictive_performance}. Because the residuals are strongly autocorrelated, each 30-day window contains far fewer independent observations than its 1440 values, so the absolute deviations overstate the evidence against the models and are used only to compare models on the same windows.} \correction{Without corrections, the deviations increase in both winters of the evaluation period and revert in spring, although with embedding this happens only at $T_\mathrm{Top}$.} Possible explanations of a winter increase include unmodelled physical processes or changes in boundary conditions that are not represented in the calibration data, such as water evaporation and condensation, frost, ice formation, or snow accumulation, and a calibrated variance that is missing a seasonal component.

	\begin{table}[htbp]
\centering
\caption{Kolmogorov-Smirnov deviation $d$ for the training month (June 2024, 30-day window centred in the month; nominal noise $\sigma_k=0.1~^\circ$C and calibrated noise).}
\label{tab:deviation_trainingmonth}
\begin{tabular}{llcccccc}
\hline
& & \multicolumn{2}{c}{Only sensors} & \multicolumn{2}{c}{Det. corrections} & \multicolumn{2}{c}{Embedded} \\
\cmidrule(lr){3-4} \cmidrule(lr){5-6} \cmidrule(lr){7-8}
Sensor & Deviation & \makecell{Nominal \\ noise} & \makecell{Calibrated \\ noise} & \makecell{Nominal \\ noise} & \makecell{Calibrated \\ noise} & \makecell{Nominal \\ noise} & \makecell{Calibrated \\ noise} \\
\hline
\multirow{2}{*}{$T_\mathrm{Bottom}$} & One-sided & 0.5476 & 0.3151 & 0.5492 & 0.3495 & 0.0177 & 0.0103 \\
 & Two-sided & 0.5441 & 0.3116 & 0.5457 & 0.3460 & 0.2589 & 0.2356 \\
\hline
\multirow{2}{*}{$T_\mathrm{Top}$} & One-sided & 0.7965 & 0.0000 & 0.8616 & 0.3048 & 0.2135 & 0.1923 \\
 & Two-sided & 0.7930 & 0.0933 & 0.8581 & 0.3013 & 0.2099 & 0.1888 \\
\hline
\multirow{2}{*}{$T_\mathrm{North}$} & One-sided & 0.5324 & 0.1255 & 0.5655 & 0.1649 & 0.0073 & 0.0000 \\
 & Two-sided & 0.5288 & 0.1649 & 0.5620 & 0.1936 & 0.2316 & 0.2404 \\
\hline
\multirow{2}{*}{$T_\mathrm{South}$} & One-sided & 0.6934 & 0.1515 & 0.6969 & 0.1636 & 0.3266 & 0.0985 \\
 & Two-sided & 0.6899 & 0.1480 & 0.6934 & 0.1601 & 0.3231 & 0.0950 \\
\hline
\end{tabular}
\end{table}

\subsection{Proper scores of the calibrated predictive distributions in and out of sample}
\label{sub:predictive_performance}
\begin{figure}[!htbp]
	\centering
	\includegraphics[width=\textwidth]{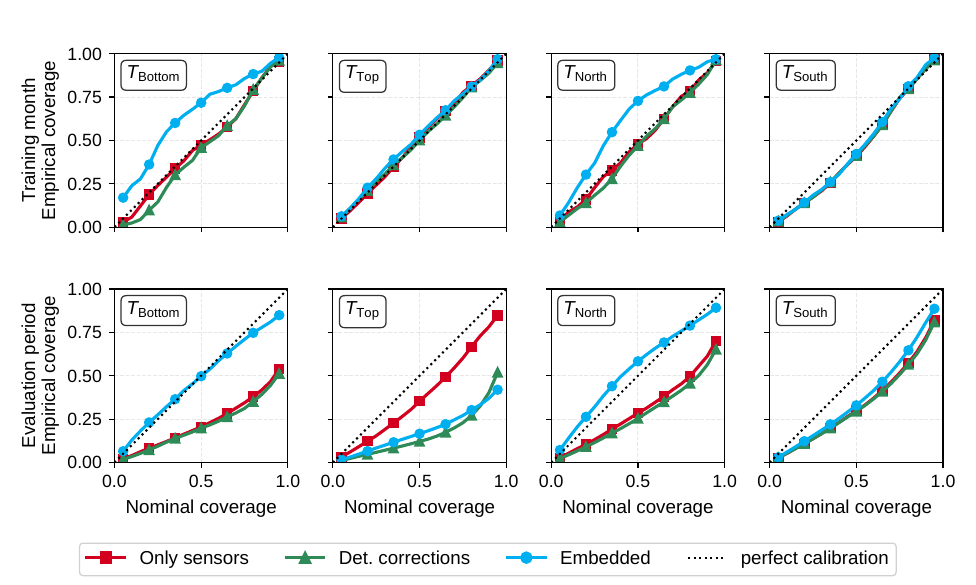}
	\caption{Reliability of the predictive distributions of the three calibrations with the noise of step~3: empirical coverage of the central predictive intervals against their nominal level, per sensor, in the training month (June 2024, top) and over the evaluation period (July 2023 to August 2025, bottom). Curves above the diagonal indicate over-dispersed, curves below under-dispersed predictive distributions.}
	\label{fig:predictive_reliability}
\end{figure}

\correction{The calibration of the embedding maximizes the logarithmic score of the predictive distribution on the training month. Whether the calibrated predictive distribution remains adequate outside this month is assessed with proper scores on the multi-year dataset, from July 2023 to August 2025, for the three calibrations with their noise of step~3, as they would be used for prediction. All metrics use the Gaussian predictive distribution of sensor $k$ at time $t$ with mean $m(t,k)$, variance $s^2(t,k)=\sigma^2_\mathrm{PCE}(t,k)+\sigma_k^2$ and standardized residual $z(t,k)=\left(y(t,k)-m(t,k)\right)/s(t,k)$. With $\Phi$ and $\varphi$ the cumulative distribution and density functions of the standard normal distribution, the empirical coverage of the central predictive interval with nominal level $p$ is}
\begin{equation}
	\hat{c}_k(p)=\frac{1}{N_t}\sum_{t=1}^{N_t}\mathbb{I}\left\{\left|z(t,k)\right|\le\Phi^{-1}\left(\tfrac{1+p}{2}\right)\right\},
	\label{eq:coverage}
\end{equation}
\correction{where $\mathbb{I}\{\cdot\}$ denotes the indicator function. Plotted against $p$ (Figure~\ref{fig:predictive_reliability}), curves above the diagonal $\hat{c}_k(p)=p$ indicate over-dispersed and curves below under-dispersed predictive distributions. The coverage assesses calibration only. A predictive distribution with the same wide spread at every time step (for example, the overall spread of the observed temperatures) can reach the nominal coverage without being informative about the individual time steps. Calibration and sharpness are assessed jointly by proper scoring rules \citep{Gneiting2007}: the continuous ranked probability score, $\mathrm{CRPS}(t,k)=\int_{-\infty}^{\infty}\left(F_{t,k}(x)-\mathbb{I}\{y(t,k)\le x\}\right)^2\mathrm{d}x$ with $F_{t,k}$ the predictive distribution function, and the logarithmic score, $\mathrm{LS}(t,k)=\ln\mathcal{N}\left(y(t,k);m(t,k),s^2(t,k)\right)$, which for a Gaussian predictive distribution take the closed forms}
\begin{equation}
	\correction{\begin{aligned}
	\mathrm{CRPS}(t,k)&= s(t,k)\left\lbrace z(t,k)\left[2\Phi\left(z(t,k)\right)-1\right]+2\varphi\left(z(t,k)\right)-\frac{1}{\sqrt{\pi}}\right\rbrace,\\
	\mathrm{LS}(t,k)&=-\frac{1}{2}\ln\left(2\pi s^2(t,k)\right)-\frac{z^2(t,k)}{2}.
	\end{aligned}}
	\label{eq:crps}
\end{equation}
\correction{The CRPS is expressed in the unit of the observations and reduces to the absolute error for a deterministic prediction, where lower values are better. Similarly, higher values of the logarithmic score are better, summed over the training data with the nominal noise, it is the log-likelihood of Eq.~\eqref{eq:log_likelihood_mfu}. Table~\ref{tab:predictive} reports both scores and the coverage of the central 90\% interval, averaged over time, at $T_\mathrm{Top}$ and over the other three sensors.}

\begin{table}[htbp]
\centering
\caption{Predictive performance of the calibrated models with the noise of step~3 at $T_\mathrm{Top}$ and averaged over $T_\mathrm{Bottom}$, $T_\mathrm{North}$ and $T_\mathrm{South}$, in the training month and over the evaluation period from July 2023 to August 2025 (all months and by season): continuous ranked probability score (CRPS, lower is better), logarithmic score (LS, higher is better) and empirical coverage of the central 90\% predictive interval. Best CRPS and LS per period and group in bold.}
\label{tab:predictive}
\setlength{\tabcolsep}{4pt}
\begin{tabular}{llcccccc}
\hline
& & \multicolumn{3}{c}{$T_\mathrm{Top}$} & \multicolumn{3}{c}{$T_\mathrm{Bottom}$, $T_\mathrm{North}$, $T_\mathrm{South}$} \\
\cmidrule(lr){3-5} \cmidrule(lr){6-8}
\textbf{Period} & \textbf{Model} & CRPS [$^\circ$C] & LS & Cov. 90\% & CRPS [$^\circ$C] & LS & Cov. 90\% \\
\hline
Training (June 2024) & Only sensors & 0.697 & -1.64 & 0.91 & 0.169 & -0.12 & 0.91 \\
 & Det. corrections & \textbf{0.464} & \textbf{-1.22} & 0.89 & \textbf{0.167} & \textbf{-0.09} & 0.91 \\
 & Embedded & 0.472 & -1.23 & 0.90 & 0.189 & -0.35 & 0.95 \\
\hline
Evaluation (all) & Only sensors & \textbf{1.042} & \textbf{-2.24} & 0.78 & 0.347 & -1.91 & 0.60 \\
 & Det. corrections & 1.429 & -4.30 & 0.40 & 0.359 & -2.16 & 0.57 \\
 & Embedded & 1.382 & -6.81 & 0.37 & \textbf{0.341} & \textbf{-1.23} & 0.82 \\
\hline
\quad winter (Dec--Feb) & Only sensors & \textbf{1.234} & \textbf{-2.45} & 0.72 & 0.497 & -3.58 & 0.39 \\
 & Det. corrections & 1.675 & -4.91 & 0.23 & 0.519 & -4.12 & 0.34 \\
 & Embedded & 1.791 & -12.81 & 0.02 & \textbf{0.453} & \textbf{-1.82} & 0.72 \\
\hline
\quad summer (Jun--Aug) & Only sensors & \textbf{0.850} & \textbf{-2.01} & 0.85 & \textbf{0.222} & -0.78 & 0.79 \\
 & Det. corrections & 1.046 & -3.36 & 0.64 & 0.224 & -0.83 & 0.78 \\
 & Embedded & 0.892 & -2.06 & 0.76 & 0.233 & \textbf{-0.65} & 0.93 \\
\hline
\end{tabular}
\end{table}

\correction{In the training month, the deterministic corrections have the best scores in both groups, and at the three sensors other than the deck top the embedded calibration is over-dispersed (Table~\ref{tab:predictive}). Over the evaluation period, the two groups give opposite results. At $T_\mathrm{Bottom}$, $T_\mathrm{North}$ and $T_\mathrm{South}$, whose discrepancy is represented by the wide spread of $\tilde c_\mathrm{ext}$, the embedded calibration has the best scores and the coverage closest to nominal (Figure~\ref{fig:predictive_reliability}). At $T_\mathrm{Top}$, it is the worst of the three, because outside the training month its predictive mean overestimates $T_\mathrm{Top}$ by 1.6~°C on average and by 2.0~°C in winter, similarly to the deterministic corrections and more than the model without corrections (1.1~°C). The offset has a clear origin. In June, the model without corrections underestimates $T_\mathrm{Top}$ by 0.7~°C, and the calibrated top factor $\tilde c_\mathrm{top}$ and $c_r$ remove this offset by warming the deck top, by a similar amount in every month of the evaluation period. The offset of the model without corrections, however, changes with the season to about $-2$~°C in winter. Because $\tilde c_\mathrm{top}$ and $c_r$ are strongly correlated, their deterministic combination absorbs the constant offset of June, leaving a narrow spread to the embedded $\tilde c_\mathrm{top}$, which does not cover the seasonal change absent from the training month.

The comparison thus shows both sides of a calibration on a single month. A discrepancy that varies in time within the training month, as at the webs and the bottom slab, is represented by the embedded spread, and the predictive distribution transfers to other seasons. A constant offset, as at the deck top, is absorbed by a deterministic correction, and its change outside the training month is not represented. The deterministic corrections, fit the training month best and transfer worst. A prediction-aware calibration can only represent the discrepancy present in its calibration data, and data covering several seasons, or a seasonal or time-dependent embedding, would be required to represent the seasonal one.}

		\subsection{Discussion on model comparison}
	A central question addressed by this work is how different model formulations should be compared and how to determine whether one model is preferable to another. Within the proposed framework, model assessment cannot be based on a single criterion. A model that achieves the smallest residual error may nevertheless provide unreliable predictive uncertainty, whereas a model with excessively large uncertainty may trivially accommodate the observations by inflating its predictive variance. %
	
	For applications involving uncertainty quantification and digital twins, the adequacy of a model should therefore also be assessed from a predictive perspective. \review{In this context, prediction should not be understood solely as the ability to reproduce the observations used for calibration, but rather as the ability to provide reliable uncertainty estimates for unobserved conditions and quantities of interest.} The evaluation should therefore consider whether the predictive distributions remain statistically consistent with available observations while providing informative uncertainty estimates for the intended prediction tasks. Such an approach has recently been explored in the context of prediction-centric uncertainty quantification \citep{Shen2025, White2026}, where models are evaluated based on their ability to produce reliable predictions for quantities relevant to downstream analyses. Model assessment may alternatively benefit from considering the target information or decisions that the predictions must support \citep{Strong2014}.
	
	In this study, model adequacy is examined through complementary diagnostics that characterize both the magnitude and the structure of predictive uncertainty, with MFU at the center. Variance decomposition provides insight into how predictive uncertainty is distributed between \correction{embedded model error, residual model error and measurement noise}, while Kolmogorov-Smirnov deviation metrics assess the agreement between predictive distributions and observed data\correction{, and proper scoring rules assess calibration and sharpness of the predictive distributions jointly \citep{Gneiting2007}. The coverage of the predictive intervals alone is not sufficient for this purpose, since it rewards over-dispersed predictions}. \review{Although these diagnostics are necessarily evaluated using observed quantities, they provide insight into whether the uncertainty model is sufficiently realistic for subsequent predictions beyond the calibration dataset.} Together, these measures allow the identification of models that simultaneously maintain realistic predictive variance and adequate distributional agreement.
	
	\section{Quantity of Interest propagation}
	\label{sec:qoi}
	\review{One potential application of the thermal response model is the thermomechanical compensation of structural deformations. As a representative QoI, we consider the thermal curvature of the bridge cross-section. Owing to the vertical temperature gradient generated by the solar heat input on the upper surface and the cross-sectional geometry, a thermal curvature, $\kappa_T$, can be computed from the temperature field. Under the assumptions of small deflections and beam kinematics, this curvature can subsequently be related to the structural deformation. In the particular case of a constant curvature along a bridge of length $L$, the corresponding vertical displacement is proportional to $\kappa_TL^2$, while $\kappa_TL$ represents the associated rotation. The thermal curvature is adopted here as the QoI, independently of the specific structural model used to infer the resulting deformation.
	
	When only temperature sensor observations are available, the curvature at a given time step is estimated as
		\begin{equation}
		\kappa^{sensor}_T=
		\beta
		\frac{T_{Top}-T_{Bottom}}
		{y_{Top}-y_{Bottom}},
	\end{equation}
	where $y_{Top}$ and $y_{Bottom}$ denote the vertical positions of the corresponding sensors and $\beta$ is the coefficient of thermal expansion. This estimate is obtained directly from the monitoring system and its uncertainty is limited to the measurement uncertainty associated with the temperature sensors. Alternatively, the calibrated thermal model, together with the embedding, enables the propagation of the MFU. In this case, the curvature is computed from the predicted temperature field by integrating the temperature distribution over the cross-section as
	\begin{equation}
		\kappa^{int}_T =
		\frac{\beta}{I}
		\int_A
		T(y)(y-y_c)dA
		\qquad
		\text{with}
		\qquad
		y_c=
		\frac{\int_A ydA}
		{\int_A dA},
		\label{eq:curvature_int}
	\end{equation}
	where $A$ is the cross-sectional area, $I$ is the second moment of area, and $y_c$ is the vertical coordinate of the centroid. Representative values of $\beta=8\times10^{-6}$ 1/K and $I=\correction{64.85}$ m$^4$ (\correction{computed from the model cross-section}) are adopted in this analysis. In practice, these quantities could also be calibrated and treated as additional sources of MFU, but they are considered deterministic here in order to isolate the uncertainty associated with the thermal model. Since $T$ depends on the embedded parameters, the statistics of the curvature, namely $\mathbb{E}[\kappa^{int}_T]$ and $\mathrm{Var}[\kappa^{int}_T]$, are obtained by propagating the embedded parameters through Equation~\ref{eq:curvature_int} using PCEs.
	
	The resulting curvatures are shown in Figure~\ref{fig:qoi}. Both approaches capture the same overall temporal evolution\correction{, but the full-field curvature differs noticeably from the sensor-based one. Over the whole month, it is smaller at 71\% of the time steps, with a mean difference of $-1.1\times10^{-6}$~1/m.} The calibrated model also provides confidence intervals that account for the uncertainty introduced during calibration, providing a quantitative measure of the reliability of the reconstructed curvature. \correction{Their median standard deviation over the first 200~h is 37\% of the root-mean-square of the sensor-based curvature and about 1.5 times that obtained when the embedded variance of $T_\mathrm{Top}$ and $T_\mathrm{Bottom}$ is propagated to the sensor-based estimate. This is because the embedding perturbs the boundary exchange on the whole external surface, including the parts of the cross-section where the temperature field is not observed. The sensor-based curvature lies within the one-standard-deviation band at 80\% of the time steps of the month, and in the first 200~h never further than 2.1 standard deviations from the predictive mean.} This uncertainty can be propagated directly to the QoI and subsequently through a structural model, enabling uncertainty-aware thermomechanical compensation.
	
	\begin{figure}[!htbp]
		\centering
		\includegraphics[width=\textwidth]{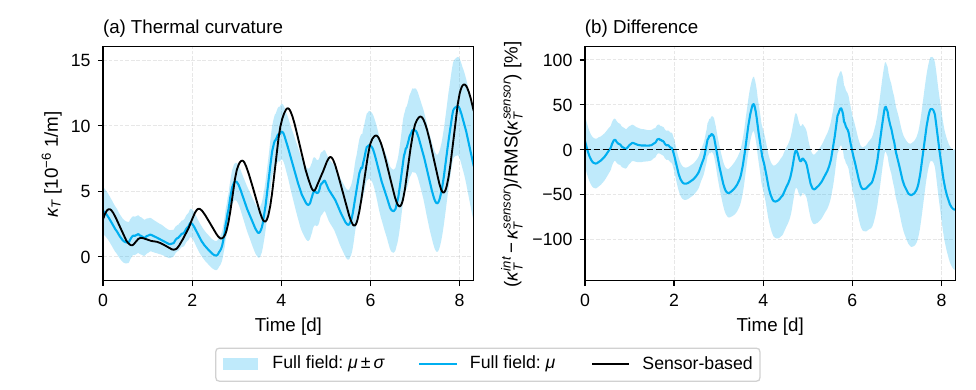}
		\caption{Curvature plots over the first 200 h of June~2024. (Left) Curvature comparison for full-field predicted curvature with confidence interval of 1$\sigma$ and the curvature predicted only from sensor observations. (Right) Difference between full-field and only-sensor curvature computations\correction{, normalized by the root-mean-square of the sensor-based curvature over the 200~h}}\label{fig:qoi}
	\end{figure}
}

	\section{Conclusions} \label{sec:conclusion}
	\review{This work developed a general probabilistic calibration framework for physics-based models in which MFU is explicitly represented through stochastic embedding of selected model parameters. The framework combines variance decomposition, distributional discrepancy metrics, and uncertainty propagation to provide a structured assessment of model adequacy and predictive reliability. Rather than treating model discrepancy solely as an additional observational noise contribution, the proposed approach separates uncertainty associated with measurements from uncertainty arising from limitations of the physical model representation. The thermal model of the Nibelungenbrücke was used as a demonstration case to investigate how explicit MFU representation affects uncertainty attribution, predictive distributions, and the propagation of uncertainty to \correction{QoIs}.
	
	\correction{The results show that explicitly representing MFU provides information that cannot be obtained from additive observational noise alone. A noise term can improve the distributional agreement with the measurements, but its uncertainty remains attached to the observations, does not modify the model response and cannot distinguish measurement variability from deficiencies of the physical model. The stochastic embedding instead attributes part of the discrepancy to the model response itself, as embedded model error. Against a deterministic calibration of the same correction factors, which isolates the effect of the embedding, the embedding reduces the noise of the deck-top sensor from 0.82 to 0.28~°C and carries most of the predictive variance at three of the four sensors, with spreads far larger than their estimation uncertainty.} \correction{The advantage of the embedded formulation also lies in the propagation of this uncertainty beyond the calibrated quantities. Through the physical model, it reaches unmeasured and derived quantities, such as the thermal curvature, in a physically consistent manner, which is essential for digital twins that predict system states beyond the sensor measurements. The main benefit of MFU-aware calibration is an interpretable uncertainty representation that can be propagated to downstream predictions. Outside the training month, the embedded calibration gives the most reliable predictive distributions at the three sensors whose discrepancy is carried by the spread of the factor of the external faces. At the deck-top sensor, the constant offset of the training month is absorbed by the deterministic deck-top correction, the seasonal offset of the model is not represented, and the predictive distribution is too narrow.}

	\correction{The analysis also shows the importance of assessing model adequacy under different operating conditions. The predictive performance of the thermal model of the Nibelungenbrücke is stable under environmental conditions comparable to those of the calibration data, whereas larger discrepancies occur in winter, indicating that the simplified two-dimensional representation captures the dominant thermal mechanisms under moderate conditions but neglects processes that become relevant in colder regimes. MFU quantification thus identifies the conditions under which simplified models remain reliable and where additional physical modelling is required.}} Several limitations of the present framework should be additionally acknowledged. First, the identifiability of embedded MFU depends on the information content of the available data and on the selected embedding structure. When only a limited number of parameters are embedded, multiple sources of model discrepancy may be aggregated into a single stochastic contribution, reducing the interpretability of the resulting uncertainty attribution. Second, the calibration in this study relies primarily on observations from a limited period, leading to seasonal variations in predictive accuracy when the model is applied throughout the year. \correction{Third, the residuals are strongly autocorrelated in time, whereas the likelihood assumes independence, so the standard errors and the absolute Kolmogorov-Smirnov deviations are overconfident and are used only for comparisons. Fourth, the common factor of the external faces makes the predictive distributions at $T_\mathrm{Bottom}$ and $T_\mathrm{North}$ over-dispersed in the training month, and at the deck top the narrow embedded spread does not cover the seasonal offset of the model. The embedded factors are hence carriers of model discrepancy and cannot be interpreted purely as convection coefficients. Fifth, the estimation uncertainty of the step~1 parameters, two of which reach the bounds of their domains, is not propagated by the staged calibration.} Finally, the simplified two-dimensional geometry neglects physical mechanisms that may influence the thermal response of the bridge, particularly under unseen environmental conditions for the calibration dataset.

	Future work will focus on extending the proposed framework towards broader operational digital-twin applications. Incorporating monitoring data from multiple seasons and operating conditions could improve the robustness of parameter inference and reduce seasonal biases in predictive performance. However, capturing the full range of environmental variability may require time-dependent representations of model parameters, which introduces additional challenges related to identifiability, calibration complexity, and uncertainty interpretation. \correction{At the deck top, embedding the deck-top factor and $c_r$ jointly, with a correlated germ, or their identifiable combination would let the calibrated spread act along the direction of their correlation.} A further extension concerns the transfer of the calibrated thermal model to a three-dimensional thermomechanical representation of the bridge, enabling uncertainty propagation from thermal predictions to structural response quantities. Such extensions will require careful assessment of additional sources of model-form uncertainty introduced by increased model complexity. Finally, integrating MFU-aware calibration into continuous digital-twin workflows would allow predictive uncertainties to be updated as new monitoring data become available \citep{Ward2021}. In such settings, a critical challenge will be distinguishing discrepancies caused by changes in the physical system, such as damage or altered boundary conditions, from discrepancies arising from model inadequacy or sensor degradation.
	
	Overall, this study demonstrates that MFU-aware probabilistic calibration provides a general framework for transforming simplified physics-based models into uncertainty-aware predictive models suitable for digital-twin environments. By distinguishing \correction{the measurement noise and the residual model error from the embedded model error, which arises from model inadequacy and can be propagated}, the proposed approach enables a more informative evaluation of predictive confidence. Its primary contribution is the ability to propagate physically interpretable uncertainty from model calibration to unmeasured quantities, thereby supporting more reliable prediction, monitoring, and decision-making in engineering applications.

\section*{Declaration of competing interest}
The authors declare that they have no known competing financial interests or personal relationships that could have appeared to influence the work reported in this paper.

\section*{Data availability}
The data that support the findings of this study are available in \url{https://doi.org/10.5281/zenodo.24687224} \citep{AndresArconesZenodo}. Monitoring data is property of Landesbetrieb Mobilität Worms (LBM Worms).

\section*{Declaration of generative AI and AI-assisted technologies in the manuscript preparation process}
\correction{During the preparation of this work the authors used ChatGPT (GPT-4 and GPT-5, OpenAI) to improve the clarity and grammar of the text, and Claude (Claude Sonnet 4.6 through GitHub Copilot, and Claude Opus 5.5 through Claude Code, Anthropic) to assist with programming the code used to obtain the results and with editing the manuscript. After using these tools, the authors reviewed and edited the content as needed and take full responsibility for the content of the published article.}

\section*{Acknowledgements}
We gratefully thank Marx Krontal Partner (MKP GmbH.) and Landesbetrieb Mobilität Worms (LBM Worms) for supplying the data of the Nibelungenbrücke. We also thank Chongjie Kang and TU Dresden for the elaboration of the cross-section drawing of Figure~\ref{fig:cross-section}.

\section*{Funding}
This work has been funded by the German Research Foundation (DFG) under grant 501811638 within the project ``C07: Data-driven Model Adaptation for Identifying Stochastic Digital Twins of Bridges'' of the priority program SPP 2388 ``Hundert plus - Verlängerung der Lebensdauer komplexer Baustrukturen durch intelligente Digitalisierung'' in Phase I and under grant 562812195 within the project ``A04: Stochastic Digital Twins of Bridges for Computing Condition Indicators under Model Form Uncertainty'' of the same program in Phase II.

\section*{Author contributions}
Conceptualization: M.W.; P.-S.K.; J.F.U. Methodology: D.A.A. Software: D.A.A. Formal analysis: D.A.A. Investigation: D.A.A. Data curation: D.A.A. Visualization: D.A.A. Writing -- original draft: D.A.A. Funding acquisition: M.W.; J.F.U. Project administration: M.W.; P.-S.K.; J.F.U. Resources: J.F.U. Supervision: M.W.; P.-S.K.; J.F.U. Writing -- review \& editing: M.W.; P.-S.K.; J.F.U.

\bibliographystyle{apalike}
\bibliography{mybibliography}

\appendix

\correction{
\section{Representation of the embedded parameters}
\label{ap:pce_order}
The predictive moments of the embedded model are computed with a tensor Gauss-Hermite quadrature in the germs $\xi_\mathrm{top}$ and $\xi_\mathrm{ext}$ of the two embedded factors (Section~\ref{sub:bayesian_calibration}). An order-$N$ rule evaluates the model at $(N+1)^2$ parameter combinations, so the cost of each evaluation of the likelihood grows quadratically with the order. To check the order-4 rule used in the calibration, the calibrated model was evaluated with rules of orders 2 to 8 (9 to 81 nodes) at the calibrated parameters (Figure~\ref{fig:pce_order}).

The log-likelihood with the nominal noise increases monotonically with the order, with decreasing increments (Figure~\ref{fig:pce_order}a). The predictive moments converge faster: the median deviation of the predictive means at order 4 from those at order 8 is at most 0.01~°C at the four sensors, and the largest deviations of the predictive standard deviations are a few hundredths of a degree (Figure~\ref{fig:pce_order}b). The deck-top factor, whose spread is small, is resolved already at low order; the remaining deviations come from the wide distribution of the factor of the external faces and reach the sensors on the webs and the bottom slab. The median standard deviation of the full-field curvature at order 4 differs by 2\% from that at order 8.

The order-4 rule is therefore adequate for the predictive moments and for the curvature. The log-likelihood is less converged than the moments, because it weights the deviations of the moments with the inverse of the nominal noise variance; a calibration with a higher order would shift the optimum, which is not quantified here. The spread of the external factor determines the required order, as a wider distribution places the outer nodes at more extreme values of the factor, where the response saturates, and requires more nodes to resolve the transition between the regimes.

\begin{figure}[!htbp]
	\centering
	\includegraphics[width=\textwidth]{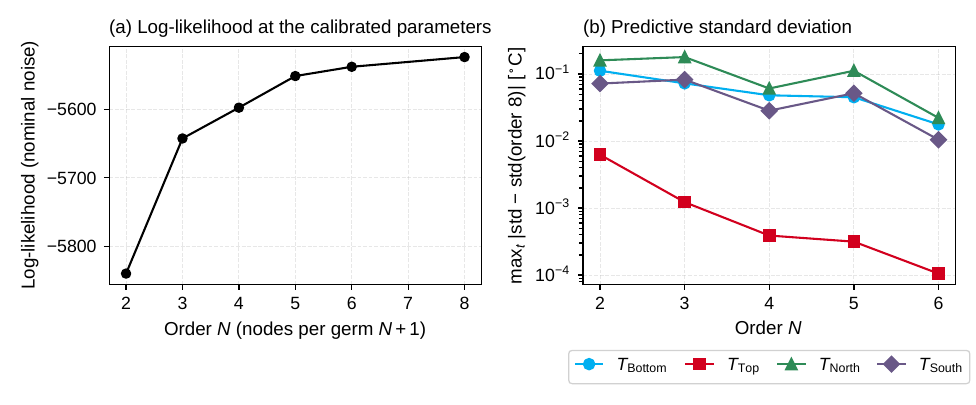}
	\caption{Convergence of the tensor Gauss-Hermite rule at the calibrated parameters of the embedded model (training month, June 2024). (a) Log-likelihood with the nominal noise $\sigma_k=0.1~^\circ$C against the order $N$ of the rule ($N+1$ nodes per germ). (b) Largest deviation over the training month of the predictive standard deviation of each sensor from that at order 8.}
	\label{fig:pce_order}
\end{figure}

}

\end{document}